\documentclass[aos]{imsart}
\RequirePackage[authoryear]{natbib}

\usepackage{amsmath,amsthm,amsfonts,amssymb,bbm,bm}
\usepackage{graphicx,wrapfig,float,enumitem,url,booktabs,caption,subcaption,comment,multirow,color,tikz,wrapfig,hyperref}
\usepackage[flushleft]{threeparttable}
\usepackage[T1]{fontenc}
\usepackage{algorithm,algpseudocode}

\makeatletter

\newcommand*{\addFileDependency}[1]{
\typeout{(#1)}
\@addtofilelist{#1}
\IfFileExists{#1}{}{\typeout{No file #1.}}
}\makeatother

\allowdisplaybreaks
\def\eop{\hfill {\large $\Box$}}
\newtheorem{theorem}{Theorem}

\newtheorem{definition}{Definition}

\newtheorem{assumption}{Assumption}

\newcommand{\bB}{\boldsymbol{B}}

\newcommand{\bI}{\boldsymbol{\mathrm{I}}}

\newcommand{\bM}{\boldsymbol{M}}

\newcommand{\bS}{\boldsymbol{S}}

\newcommand{\bU}{\boldsymbol{U}}

\newcommand{\bV}{\boldsymbol{V}}

\newcommand{\bW}{\boldsymbol{W}}

\newcommand{\bX}{\boldsymbol{X}}
\newcommand{\bx}{\boldsymbol{x}}
\newcommand{\bY}{\boldsymbol{Y}}
\newcommand{\by}{\boldsymbol{y}}

\newcommand{\bzero}{\boldsymbol{0}}
\newcommand{\balpha}{\boldsymbol{\alpha}}
\newcommand{\bbeta}{\boldsymbol{\beta}}
\newcommand{\bdelta}{\boldsymbol{\delta}}
\newcommand{\bepsilon}{\boldsymbol{\epsilon}}
\newcommand{\bGamma}{\boldsymbol{\Gamma}}
\newcommand{\bgamma}{\boldsymbol{\gamma}}
\newcommand{\bmu}{\boldsymbol{\mu}}

\newcommand{\bSigma}{\boldsymbol{\Sigma}}

\newcommand{\btheta}{\boldsymbol{\theta}}
\newcommand{\bTheta}{\boldsymbol{\Theta}}
\newcommand{\bzeta}{\boldsymbol{\zeta}}
\newcommand{\bkappa}{\boldsymbol{\kappa}}
\newcommand{\bLambda}{\boldsymbol{\Lambda}}

\newcommand{\Ebb}{\mathbb{E}}
\newcommand{\Pbb}{\mathbb{P}}
\newcommand{\Rbb}{\mathbb{R}}
\newcommand{\Pa}{\mathrm{Pa}}
\newcommand{\vh}{\text{vech}}

\newcommand{\trans}{^{\mbox{\tiny{\sf  T}}}}
\newcommand{\indep}{\;\, \rule[0em]{.03em}{.67em} \hspace{-.27em}
	\rule[-.02em]{.7em}{.03em} \hspace{-.27em}
	\rule[0em]{.03em}{.67em}\;\,}

\usepackage{xr}
\begin{document}

\begin{frontmatter}

\title{Multivariate Dynamic Mediation Analysis under a Reinforcement Learning Framework} 

\begin{aug}
\author[A]{\fnms{Lan}~\snm{Luo}\ead[label=e1]{l.luo@rutgers.edu}}\footnote[1]{Equal contribution.},
\author[B]{\fnms{Chengchun}~\snm{Shi}\ead[label=e2]{C.Shi7@lse.ac.uk}}\footnotemark,
\author[C]{\fnms{Jitao}~\snm{Wang}\ead[label=e3]{jitwang@umich.edu}}${\color{blue}{}^*}$,
\author[C]{\fnms{Zhenke}~\snm{Wu}\ead[label=e4]{zhenkewu@umich.edu}}
\and
\author[D]{\fnms{Lexin}~\snm{Li}\ead[label=e5]{lexinli@berkeley.edu}}

\address[A]{Department of Biostatistics and Epidemiology, Rutgers University \printead[presep={,\ }]{e1}}	

\address[B]{Department of Statistics, London School of Economics and Political Science\printead[presep={,\ }]{e2}}
	
\address[C]{Department of Biostatistics, University of Michigan \printead[presep={,\ }]{e3,e4}}	
	
\address[D]{Division of Biostatistics,
University of California at Berkeley \printead[presep={,\ }]{e5}}	
\end{aug}

\begin{abstract}
Mediation analysis is an important analytic tool commonly used in a broad range of scientific applications. In this article, we study the problem of mediation analysis when there are multivariate and conditionally dependent mediators, and when the variables are observed over multiple time points. The problem is challenging, because the effect of a mediator involves not only the path from the treatment to this mediator itself at the current time point, but also all possible paths pointed to this mediator from its upstream mediators, as well as the carryover effects from all previous time points. We propose a novel multivariate dynamic mediation analysis approach. Drawing inspiration from the Markov decision process model that is frequently employed in reinforcement learning, we introduce a Markov mediation process paired with a system of time-varying linear structural equation models to formulate the problem. We then formally define the individual mediation effect, built upon the idea of simultaneous interventions and intervention calculus. We next derive the closed-form expression, propose an iterative estimation procedure under the Markov mediation process model, {and develop a bootstrap method to infer the individual mediation effect}. We study both the asymptotic property and the empirical performance of the proposed methodology, and further illustrate its usefulness with a mobile health application. 
\end{abstract}



\end{frontmatter}

\section{Introduction}
\label{sec:introduction}

Mediation analysis is an important analytic tool, which seeks to explain the mechanism or pathway that underlies an observed relationship between a treatment and an outcome variable, through the inclusion of an intermediary variable known as a mediator. It decomposes the effect of the treatment on the outcome into a direct effect and an indirect effect, the latter of which indicates whether the mediator is on a pathway from the treatment to the outcome \citep{Baron1986}. Mediation analysis is widely employed in a range of scientific applications, including psychology \citep{MacKinnon2008, rucker2011mediation}, genomics \citep{huang2016hypothesis, bi2017genome}, economics \citep{celli2022causal}, social science \citep{kaufman2001epidemiology}, neuroscience \citep{zhao2022pathway}, among many others. See \citet{VanderWeele2016} for a comprehensive review and the references therein. 

Mediation analysis has seen considerable progress in recent years. There are two lines of research of particular interest, mediation analysis with multivariate mediators, and mediation analysis with time-varying variables. The first line targets the scenario where there are multiple mediators, and one central goal is to evaluate and quantify the contribution of the treatment on the outcome attributed to each individual mediator. There are three main categories of solutions. One category explicitly imposes that the multivariate mediators are conditionally independent given the treatment, which substantially simplifies the analysis \citep{boca2014testing, huang2016hypothesis, zhang2016estimating, guo2023statistical, shuai2023mediation,yuan2023confounding}. Another category does not impose such a condition, but instead marginalizes each individual mediator, which in effect neglects any potential interaction and dependency among the mediators \citep{sampson2018fwer, djordjilovic2022optimal, zhao2022multimodal, zhao2022pathway}. The last category does allow correlated mediators, characterizes their dependency through some unknown directed acyclic graph (DAG), then carries out mediation analysis based on the estimated DAG \citep{Maathuis2009IDA, chakrabortty2018inference,cai2021anoce, shi2022testing,wei2024efficient}. The second line targets the scenario where the treatment, mediator and outcome are observed over multiple time points, or stages. There have been some pioneering works along this line \citep{selig2009mediation,preacher2015advances, vanderweele2017mediation, Lin2017-SIM, Huang2017-BDM, zheng2017longitudinal, zhao2018functional, hejazi2020nonparametric, cai2022estimation, diaz2022efficient, ge2023reinforcement}. Nevertheless, they all focus on the case where there is only a single mediator, and most only consider the case where the number of time points, or the time horizon, is finite. 

In this article, we study the problem of mediation analysis when there are multivariate and conditionally dependent mediators, and when the variables are observed over multiple time points. Our motivation is a mobile health application from the Intern Health Study \cite[IHS,][]{necamp2020assessing}. It is a 26-week prospective longitudinal randomized trial targeting first-year training physicians in the United States \citep{necamp2020assessing}. A key objective of this study is to investigate the effectiveness of in-the-moment mobile prompts in improving an intern's mood score while minimizing user burden and expense. The prompts, delivered through a customized mobile app, consist of practical tips and life insights, such as reminders to have a break, take a walk, or prioritize sleep, that aim at promoting healthy behaviors among interns. These prompts not only affect an intern's mood, but also influence other physiological measurements, including physical activity, sleep duration, and heart rate variability, which are closely associated with an individual's mental state. We can naturally formulate this problem in the framework of mediation analysis, where the treatment is the binary variable that encodes receiving a prompt or not, regardless of the prompt type, the mediators consist of measurements of physical activities, sleep duration, and heart rate variability, and the outcome is the mood score. Given the presence of multiple mediators and the longitudinal nature of the data, it calls for a mediation analysis approach that tackles both multivariate mediators and multiple time points. 

However, the problem is challenging, for several reasons. First, multivariate and conditionally correlated mediators introduce considerable complexity. Unlike the single mediator or conditionally independent mediator case, the indirect effect of each individual mediator involves not only the path from the treatment to this mediator itself, but also all possible paths pointed to this mediator from its upstream mediators. The total number of potential paths that go through any mediator is super-exponential in the number of mediators. It is thus crucial to carefully disentangle the individual mediation effect when the multivariate mediators are interrelated with unknown dependency. Second, time-varying variables introduce another layer of complexity, as there are carryover effects along time. Unlike the single-time-point case, the mediator at a given time point could affect both the current and future outcomes, and thus it is crucial to learn both the immediate effect and the delayed effect. Infinite time horizon further complicates the analysis when the mediator effect accumulates over infinite time. 

To address those challenges, we propose a novel multivariate dynamic mediation analysis approach. Our proposal consists of four key components. First, drawing inspiration from the Markov decision process \citep[MDP,][]{puterman2014markov} model that is frequently employed in reinforcement learning \citep[RL,][]{sutton2018reinforcement} to handle the carryover effects over time, we introduce a Markov mediation process (MMP) framework to formulate the dynamic mediation analysis problem. We then introduce a system of time-varying linear structural equation models (SEMs) to specifically characterize the relations among the treatment, mediator and outcome variables. Second, within the framework of MMP, we formally define the \emph{individual mediation effect}, which is built upon the idea of simultaneous interventions and intervention calculus \citep{pearl2000do}. This individual mediation effect can be further decomposed into a sum of the \emph{immediate effect} and the \emph{delayed effect}, the latter of which quantifies the carryover effects of the past treatments and mediators. Third, under the proposed MMP and the linear SEMs, we derive the closed-form expression for the individual mediation effect; see Theorems \ref{thm:eta_finite} and \ref{thm:eta_infinite}. These theorems form the basis of our proposed procedure, allow us to express the individual mediation effect using a set of within-stage and cross-stage intermediate quantities that can be estimated through some recursive formulations, and consist of the main contributions of our proposal. We next propose to estimate those within-stage quantities through linear regressions with backdoor covariate adjustment, and estimate those cross-stage quantities through the transition equations under SEMs. Finally, we study the asymptotic property of our proposed estimator of the individual mediation effect, develop a bootstrap method to construct its confidence interval (CI), and conduct extensive numerical studies to demonstrate the effectiveness of our method. In summary, our approach allows us to separately evaluate the contribution of each individual mediator, where multiple mediators are interrelated with unknown dependency and are observed over multiple time points.  

The rest of the article is organized as follows. We present the Markov mediation process, the structural equation models, and the definition of the individual mediation effect in Section~\ref{sec:model}. We derive the intermediate quantities, the recursive formulation of the individual mediation effect, and the estimation algorithms in Section~\ref{sec:method}. We establish the asymptotic properties in Section~\ref{sec:theory}. We carry out the simulations in Section~\ref{sec:simulation}, and revisit the IHS example in Section~\ref{sec:real}. We relegate all technical proofs to the Supplementary Appendix.

\section{Model and Definition}
\label{sec:model}

In this section, we first present our model setup, then formally define the mediation effect of interest in the multivariate dynamic mediation analysis setting.

\subsection{Markov mediation process and structural equation models}
\label{ssec:MMP}

\begin{figure}[t!]
\centering
\includegraphics[width=0.95\linewidth, height=1.5in]{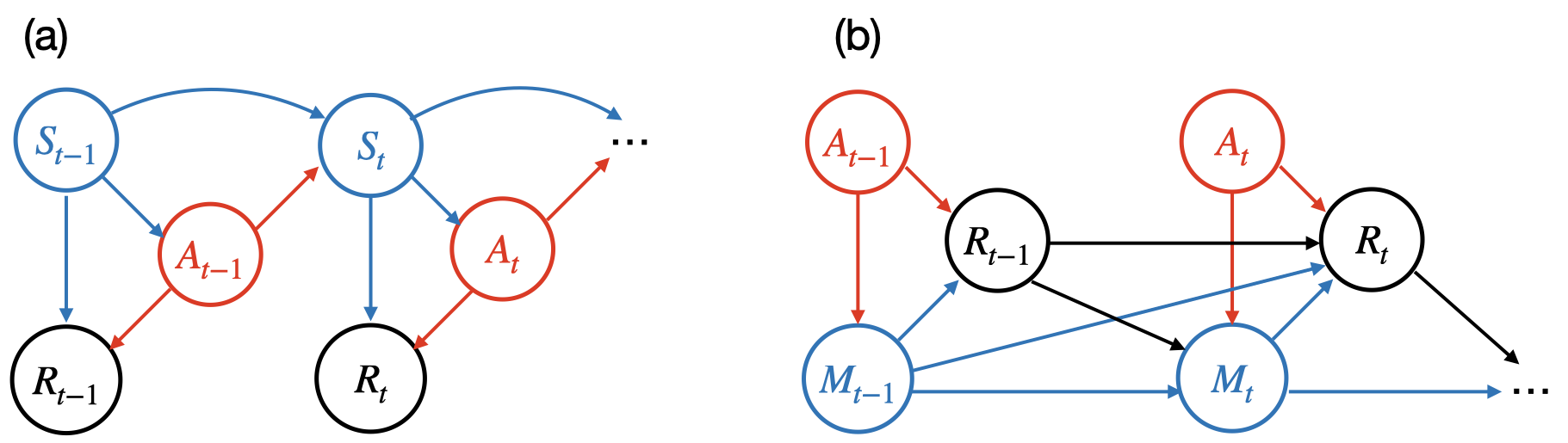}
\caption{(a) Diagram of Markov decision process (MDP), where treatments depend on current states only, and $(S_{t}, A_t, R_t)$ represents the state-treatment-reward triplet; (b) Diagram of the proposed Markov mediation process (MMP), where current mediators depend on previous mediator-reward pairs, and $(A_t, M_t, R_t)$ represents the treatment-mediator-outcome triplet.}
\label{fig:DAG}
\end{figure}

We first introduce a Markov mediation process framework to formulate the problem we target; see Figure \ref{fig:DAG}(b) for a graphical illustration. Consider the treatment-mediator-outcome triplets $\{ \left( A_t, \bM_t, R_t \right) : t = 1, \ldots, T\}$ over time, where $T$ denotes the number of time points or stages. 
At each time point or stage $t$, a random treatment $A_t \in \Rbb$ is administered, which subsequently affects a $d$-dimensional vector of potential mediators $\bM_t \in \Rbb^{d}$, and an outcome variable $R_t \in \Rbb$. We assume they satisfy the Markov assumption, in that 
\begin{equation} \label{eq:markov}
(R_t,\bM_t) \ \indep \ (R_s,\bM_{s}, A_{s+1})_{s<t-1} \ \big | \ (R_{t-1},\bM_{t-1},A_t), \; \text{ for any } t \geq 1,
\end{equation} 
where $\indep$ denotes statistical independence. We remark the Markov condition like \eqref{eq:markov} is widely imposed in sequential data problems \citep{sutton2018reinforcement}. Suppose the observed data consists of $n$ independent and identically distributed (i.i.d.) realizations of the triplets $\{(A_t,\bM_t,R_t) : t = 1, \ldots, T\}$. {We allow the data to be either densely or sparsely observed. Moreover, for simplicity, we assume that all subjects have the same $T$. Nevertheless, our proposed method can be adapted to accommodate the setting when $T$ varies among subjects, and when the time lags between two time points differ. See Section \ref{ssec:varyingT} for more details.}

In addition to the Markov condition, we further assume that the random treatment assignment is independent of the prior information, in that
\begin{equation} \label{eq:rdmtrt}
A_t \ \indep \ (A_s, \bM_s, R_s), \; \text{ for any } s < t. 
\end{equation} 
Condition \eqref{eq:rdmtrt} holds in the sequentially randomized trials naturally, including our IHS example, which is the main setting we target in this article. {Meanwhile, our framework can also be extended to the setting that involves certain static baseline confounders that may influence the treatment selection, thus accommodating the data from observational studies. Specifically, letting $X$ denote the baseline confounders, Condition \eqref{eq:rdmtrt} can be relaxed to $A_t \ \indep \ (A_s, \bM_s, R_s)|X$, for any $s<t$. The individual mediation effect that we define later can be similarly derived by incorporating $X$ into the model. For presentation simplicity, however, we choose not to include those confounders in this article.}

We next introduce a system of time-varying structural equation models, 
\begin{equation} \label{eq:LSEM}
(\bM_{t} - \bmu_t) = \bW_t\trans (\bM_{t}-\bmu_t) +\bepsilon_{wt}, 
\end{equation}
where $\bmu_t \equiv \Ebb\left( \bM_t\mid A_t, \bM_{t-1}, R_{t-1} \right)$ is the condition mean function, $\bW_{t} \in \Rbb^{d\times d}$ is the weight matrix, such that $(\bW_t)_{ij} \neq 0$ if and only if mediator $M_{ti}$ is a parent of $M_{tj}$, i.e., $M_{ti}$ is in the parent set, $M_{ti} \in \Pa(M_{tj}) \equiv \big\{ M_{ti}: (\bW_{t})_{ij} \neq 0, i \in \{1, \ldots, d\}\setminus \{j\} \big\}$, and  $\bepsilon_{wt} = (\epsilon_{w1t},\ldots,\epsilon_{wdt})\trans \in \Rbb^{d}$ is a vector of mean zero random errors. {In model \eqref{eq:LSEM}, the weight matrix $\bW_t$ models the interactions among the mediators $\bM_t$ at each time point}, and the conditional mean $\bmu_t$ characterizes the dynamic dependence over time. 

We then consider the linear models for $\bmu_t$ and $R_t$, in that
\begin{align} \label{eq:model_setup}
\begin{split}
\bmu_t & = \balpha_{1t} + \bm{\delta}_{1t} A_t + \bGamma_{1t}\bM_{t-1} + \bm{\zeta}_{1t} R_{t-1}, \\
R_t & = \alpha_{2t} + \delta_{2t} A_t + \bgamma_{2t}\trans \bM_{t-1} + \zeta_{2t} R_{t-1} +  \bkappa_t\trans \bM_{t} + \epsilon_{rt},
\end{split}
\end{align}
for some $\balpha_{1t},\bdelta_{1t},\bzeta_{1t}\in\mathbb{R}^d$, $\bGamma_{1t}\in\mathbb{R}^{d\times d}$, and for $\bgamma_{2t},\bkappa_t\in\mathbb{R}^d$, $\alpha_{2t},\delta_{2t},\zeta_{2t}\in\Rbb$, and some mean zero errors $\{\epsilon_{rt}\}_t$ independent over time, respectively. {In model \eqref{eq:model_setup}, $\bGamma_{1t}$ characterizes the effect of $\bm{M}_{t-1}$ on $\bm{M}_t$.} {All random variables in \eqref{eq:model_setup} are assumed to have finite second moments}. Let $\bTheta_{1t} \equiv (\balpha_{1t},\bdelta_{1t},\bGamma_{1t},\bzeta_{1t}) \in \Rbb^{d\times (d+3)}$ collect all the parameters for $\bmu_t$, and $\bTheta_{2t} \equiv (\alpha_{2t}, \delta_{2t}, \bgamma_{2t}\trans$, $\zeta_{2t}, \bkappa_t\trans)\trans \in \Rbb^{(2d+3)\times 1}$ collect all the parameters for $R_t$. 

We make a few remarks. First, we draw a connection between the proposed MMP and MDP commonly studied in RL -- a powerful machine learning technique for optimal sequential decision making \citep{murphy2003optimal,mnih2015human,silver2017mastering,qin2021reinforcement,zhang2023adaptive}. Both capture the carryover effects over time. MDP achieves this by introducing a sequence of time-varying feature variables, referred to as the states. It then models the carryover effects through state transitions, allowing past treatments to affect future outcomes through their impact on future states; see Figure \ref{fig:DAG}(a) for an illustration. This approach has gained substantial attention for policy evaluation, serving to model both immediate and long-term effects of a target policy  \citep{luckett2020estimating,hao2021bootstrapping, liao2021off, kallus2022efficiently, hu2022switchback, liao2022batch, ramprasad2022online, shi2022statistical,liu2023online,wang2023projected}. In a similar vein, our MMP operates by modeling the indirect influence of a preceding mediator via the transitions of mediator-outcome pairs. In essence, a past mediator influences future outcomes by exerting its effects on both the prior outcome and the subsequent mediator; see Figure~\ref{fig:DAG}(b) for an illustration. Second, the relation in \eqref{eq:LSEM} should be understood as a data generating mechanism, rather than as a mere association. It corresponds to a directed acyclic graph (DAG). Third, to keep the presentation simple, we do not include any time-varying confounders, which may be incorporated into our solution in a relatively straightforward fashion. We do not consider any unobserved confounders either, because we consider random treatment assignments. 
We leave the case with unobserved confounders as future research.  
Finally, we consider linear type models in both \eqref{eq:LSEM} and \eqref{eq:model_setup}. The analysis of mediation has been dominated by linear regression paradigms and such linearity assumption is commonly adopted in existing work, see for example,~\cite{nandy2017joint,chakrabortty2018inference,shi2022testing}. It is possible to extend to nonlinear type models, under which the definition of individual mediation effect we give later still holds, but is more difficult to evaluate. 

{In this article, we consider two different settings: the finite-horizon and the infinite-horizon, borrowing the concepts from the RL literature \citep{sutton2018reinforcement}. For the finite-horizon setting, the number of time points $T$ is fixed and finite. It usually corresponds to the traditional longitudinal studies, where subjects receive a finite number of treatments over a predetermined time period. For the infinite-horizon setting, $T$ approaches infinity theoretically. It often applies to the applications involving continuous monitoring systems such as wearable devices, where the data is densely collected at high frequencies, e.g., every second or minute. For the finite-horizon setting, we allow the process $\{ A_t,\bM_t,R_t \}$ to be non-stationary. However, for the infinite-horizon setting, we require the process to be stationary over time; that is, the parameters $\bW_t$, $\bTheta_{1t}$ and $\bTheta_{2t}$ are constant over time. Without stationarity, we need to extrapolate to infer system dynamics beyond the observed time.} 

In summary, we believe that our model framework provides a reasonable starting point for multivariate dynamic mediation analysis. The methodology under our setting is already complex enough, and deserves a careful investigation.

\subsection{Total, direct and indirect effects}
\label{ssec:totaleffect}

{Before we formally define the individual mediation effect, which is the main target of interest in this article, we first introduce the notions of \emph{total effect}, \emph{natural direct effect}, and \emph{natural indirect effect} in the finite-horizon setting. This is to facilitate the understanding of our target, and also to establish the connection with the classical mediation analysis literature. 

For the total effect, consider a hypothetical intervention applied to the system, where we set all treatments to some value $a$ uniformly over the entire population. This can be realized through Pearl's do-operator, $do(A_t=a)$, which generates an interventional distribution by removing the edges leading into $\{A_s:s\leq t\}$ in the corresponding DAG \citep{pearl2000do}. We denote the post-interventional expectation of $R_t$ by $\Ebb[R_t \mid do(A_s=a,\forall s\leq t)]$. We then define the total effect over $T$ time points as
\begin{eqnarray}\label{eqn:totaleffect}
\textrm{TE}=\sum_{t=1}^{T}\frac{\partial}{\partial a} \Ebb[R_t\mid do(A_s=a,\forall s\leq t)].
\end{eqnarray}
We make two remarks. First, each summand $\partial \Ebb[R_t\mid do(A_s=a, \forall s \leq t)]/\partial a$ on the right-hand-side of \eqref{eqn:totaleffect} measures the total effect of treatment $\{A_s\}_{s\leq t}$ on $R_t$ at time $t$. Second, when the treatment takes a discrete value, the derivative in those terms can be substituted by the difference operator. For instance, if the treatment is binary, we obtain that 
\begin{eqnarray*}
\textrm{TE}=\sum_{t=1}^{T} \Big\{\Ebb[R_t\mid do(A_s=1,\forall s\leq t)]-\Ebb[R_t\mid do(A_s=0,\forall s\leq t)]\Big\}.
\end{eqnarray*}

Next, we decompose the total effect into the sum of the natural direct effect and the natural indirect effect. Specifically, at each time point $t = 1, \ldots, T$, we define the direct effect to be the portion of the total effect of a sequence of treatment variables $\{A_s:s\leq t\}$ on the outcome $R_t$ that does not go through $\{\bM_{s} : s \leq t\}$. To measure such an effect, we consider the joint intervention on $(A_s,\bM_{s})$ through $do(A_s=a, \bM_{s}=\bm{m}_s)$, and denote the post-interventional expectation of $R_t$ by $\Ebb[R_t\mid do(A_s=a, \bM_{s}=\bm{m}_s, \forall s\leq t)]$. We then define the natural direct effect over $T$ time points as
\begin{eqnarray*}
\textrm{DE}=\sum_{t=1}^T \int_{\bm{m}_1,\cdots,\bm{m}_T} \frac{\partial}{\partial a} \Ebb[R_t\mid do(A_s=a,\bM_s=\bm{m}_s,\forall s\leq t)]\\\times f(\bm{m}_1,\cdots,\bm{m}_T|do(A_t=a,\forall 1\leq t\leq T))d\bm{m}_1\cdots d\bm{m}_T,
\end{eqnarray*}
where $f$ denotes the interventional probability density function of $(\bM_1,\cdots,\bM_T)$ under the assumption that all treatments are set to $a$, and the term $\partial \Ebb[R_t\mid do(A_s=a,\bM_s=\bm{m}_s,\forall s\leq t)]/\partial a$ corresponds to the interventional effect of $\{A_s\}_{s\leq t}$ on $R_t$ when setting the interventional values of $\{\bM_s\}_{s\leq t}$ to  constants. 

We then define the natural indirect effect as the difference between TE and DE, i.e., $\textrm{IE} = \textrm{TE} - \textrm{DE}$, which quantifies the portion of the effect of the treatment sequence on the outcome that goes through the mediator sequence.

We remark that, the definitions of TE, DE and IE do \emph{not} require linear structural equation models like \eqref{eq:model_setup}. However, imposing such models greatly simplify the analysis. More specifically, in a general Markov mediation process, both $\partial \Ebb[R_t\mid do(A_s=a, \forall s \leq t)]/\partial a$ and $\partial \Ebb[R_t\mid do(A_s=a,\bM_s=\bm{m}_s,\forall s\leq t)]/\partial a$ can depend on $(a,\bm{m}_1,\cdots,\bm{m}_t)$ in a very complex manner, as also mentioned in~\citet{chakrabortty2018inference}. However, \eqref{eq:model_setup} simplifies these expressions to be  functions of the regression coefficients that are independent of specific values of $\bm{m}_s$ or $a$; see Section \ref{sec:method} for more details. As a result, DE is equivalent to
\begin{eqnarray}\label{eqn:controlledDE}
    \sum_{t=1}^T \frac{\partial}{\partial a} \Ebb[R_t\mid do(A_s=a,\bM_s=\bm{m},\forall s\leq t)],
\end{eqnarray}
for any $\bm{m}$, where \eqref{eqn:controlledDE} is the \textit{controlled direct effect}, computed by setting all values of mediators to $\bm{m}$. Correspondingly, by \eqref{eqn:totaleffect} and \eqref{eqn:controlledDE}, the natural indirect effect becomes
\begin{eqnarray}\label{eqn:indirecteffect}
\sum_{t=1}^{T}\Big\{\frac{\partial}{\partial a} \Ebb[R_t\mid do(A_s=a,\forall s\leq t)]- \frac{\partial}{\partial a} \Ebb[R_t\mid do(A_s=a,\bM_s=\bm{m},\forall s\leq t)]\Big\}.
\end{eqnarray}
In other words, the linear structural equation model avoids the need to estimate the interventional probability density function $f$, thus substantially simplifying the forms of DE, IE, and their subsequent estimation and inference procedures. 
}

\subsection{Individual mediation effect}
\label{ssec:definition}

We now formally define the individual mediation effect, the main target in our dynamic mediation analysis. We begin with the finite-horizon setting. 

\begin{definition}[Individual mediation effect in finite-horizon]\label{def:finite}
In a finite-horizon setting, the individual mediation effect of the $j$th mediator over $T$ time points is defined as
\begin{equation*}
{\eta_{j}^{(T)} \equiv \sum_{t=1}^{T}\frac{\partial}{\partial a} \Ebb[R_t\mid do(A_s=a,\forall s\leq t)] - \sum_{t=1}^{T}\frac{\partial}{\partial a} \Ebb[ R_t\mid do(A_s=a, M_{sj}=m, \forall s\leq t)].}
\end{equation*}
\end{definition}

When the treatment is binary, $\eta_j^{(T)}$ can be defined as 
\begin{align}\label{eqn:etajT}
\begin{split}
\eta_{j}^{(T)} = \sum_{a=0}^1 (-1)^{a+1} \bigg\{\sum_{t=1}^{T}\Ebb[R_t & \mid do(A_s=a,\forall s\leq t)] \\
& - \sum_{t=1}^{T}\Ebb[R_t\mid do(A_s=a, M_{sj}=m, \forall s\leq t)]\bigg\}.
\end{split}
\end{align}

{We again make a few remarks. First, our definition of the individual mediation effect is related to the definition of IE in \eqref{eqn:indirecteffect}, and is defined as the difference between the total effect and the interventional effect $\partial \Ebb[ R_t\mid do(A_s=a, M_{sj}=m, \forall s\leq t)]/\partial_a$ over time. However, the key difference is that, whereas IE measures the mediation effect of \emph{all} mediators, $\eta_{j}^{(T)}$ focuses on quantifying the portion of the effect that specifically passes through the \emph{individual} $j$th mediator. As such, we only intervene the $j$th mediator as opposed to all mediators in the intervention, and we refer to it as an \emph{individual} mediation effect. 

Second, our definition is consistent with the existing literature. Specifically, when there is only one mediator $M$ and a single time point, our definition aligns with the classical definition of \citet{Baron1986}, ${\partial}\mathbb{E}[R\mid do(A=a)]/{\partial a} - {\partial}\mathbb{E}[R\mid do(A=a, M=m)]/{\partial a}$, where the first term is the regression coefficient by regressing $R$ on $A$, and the second term is the partial regression coefficient by further including $M$ in the regression. This difference measuring the reduction in the total effect due to controlling for $M$ is widely used to quantify the effect mediated through $M$; see also \cite{pearl2012causal}. When there is a single time point, our definition is consistent with the one proposed by \citet{chakrabortty2018inference}. When there is a single mediator, \eqref{eqn:etajT} is similar to the definitions proposed by \citet{vanderweele2017mediation} and \citet{ge2023reinforcement} under a potential outcome framework.

Third, under the linear structural equation model \eqref{eq:model_setup}, $\eta_j^{(T)}$ is a constant function with respect to $m$ and $a$. It can depend on $(a,m)$ in a more general Markov mediation process, e.g., when either the transition model from a given time point to the next or the DAG structural equation model at a given time point is nonlinear.} In addition, we observe that, under \eqref{eq:model_setup}, the second post-interventional expectation term, $\Ebb[R_t\mid do(A_s=a, M_{sj} = m, \forall s\leq t)]$, may be analyzed by the path method \citep{Wright1921path}. That is, it can be calculated by summing up the effects along all directed paths from $\{A_s\}_{s\leq t}$ to $R_t$ that do not pass through $\{M_{sj}\}_{s\leq t}$. However, this approach can be computationally expensive, since the number of paths grows exponentially fast as the number of time points $T$ increases.

{Finally, $\eta_{j}^{(T)}$ in Definition \ref{def:finite} measures the \emph{cumulative} effect mediated through the $j$th mediator $M_j$ across all $T$ stages. Alternatively, one may be interested in the \emph{incremental} effect, 
\begin{equation*}
\Delta_j^{(t)} \equiv \frac{\partial}{\partial a} \Ebb[R_t\mid do(A_s=a,\forall s\leq t)] - \frac{\partial}{\partial a} \Ebb[R_t\mid do(A_s=a, M_{sj}=m, \forall s\leq t)],
\end{equation*}
which measures the individual mediation effect attributed to $M_j$ at time $t$.  By definition, we see that this incremental effect is related to the cumulative effect, in that $\eta_j^{(T)}=\sum_{t=1}^T \Delta_j^{(t)}$.} 

To better understand our definition of the individual mediation effect, we further decompose $\Delta_j^{(t)}$ into the sum of the \emph{immediate} individual mediation effect (IIME) and the \emph{delayed} individual mediation effect (DIME), as defined next.  

\begin{definition}[Immediate and delayed individual mediation effects in finite-horizon]\label{def:DIME_IIME}
Define the immediate individual mediation effect (IIME) and the delayed individual mediation effect (DIME) as, 
\begin{eqnarray*}
\mathrm{IIME}_j^{(t)}  & \equiv & \frac{\partial}{\partial a}\Ebb[R_t\mid do(A_t=a)] - \frac{\partial}{\partial a}\Ebb[R_t\mid do(A_t=a, M_{tj}=m)], \\
\mathrm{DIME}_j^{(t)} & \equiv & \Delta_j^{(t)}-\mathrm{IIME}_j^{(t)}. 
\end{eqnarray*}
\end{definition}

\noindent
At a given time point $t$, $\text{IIME}_j^{(t)}$ can be interpreted as the change in the total effect of $A_t$ on $R_t$ when $M_{tj}$ is knocked out, whereas $\text{DIME}_j^{(t)}$ captures the individual mediation effect of the $j$th mediator that is carried over from all previous stages $s<t$ up to time $t$.
 
We next turn to the infinite-horizon setting. {In this setting, the individual mediation effect is defined as the \emph{average} effect over time, so to prevent it from being unbounded.}

\begin{definition}[Individual mediation effect in infinite-horizon]\label{def:infinite}
In an infinite-horizon setting, the individual mediation effect of the $j$th mediator is defined as $\eta_j^{(\infty)} \equiv \underset{T\to\infty}{\lim} \eta_j^{(T)}/T$, provided that the limit exists.
\end{definition}

We comment that all above definitions are natural and agree with the intuitions. We next discuss how to evaluate and estimate the individual mediation effect given the data.

\section{Evaluation of Individual Mediation Effect}
\label{sec:method}

In this section, we propose approaches to estimate and infer the individual mediation effects given in Definitions~\ref{def:finite} and \ref{def:infinite}. The problem, however, is very challenging, as we have to deal with both the multiple mediators with unknown correlation structure, as well as the carryover effects from the upstream treatments and mediators. Our proposed solution involves deriving a closed-form expression for the individual mediation effect, as detailed in Theorems \ref{thm:eta_finite} and \ref{thm:eta_infinite}. This expression is dependent on several intermediate quantities that are computable through a set of recursive relations. Leveraging these theoretical findings, we develop an efficient recursive computation algorithm to estimate these intermediate quantities, which subsequently facilitate both the estimation and inference of the individual mediation effect.

\subsection{An illustrative example} 
\label{ssec:two-stage}

\begin{figure}[t!]
\includegraphics[width=0.7\linewidth, height=1.85in]{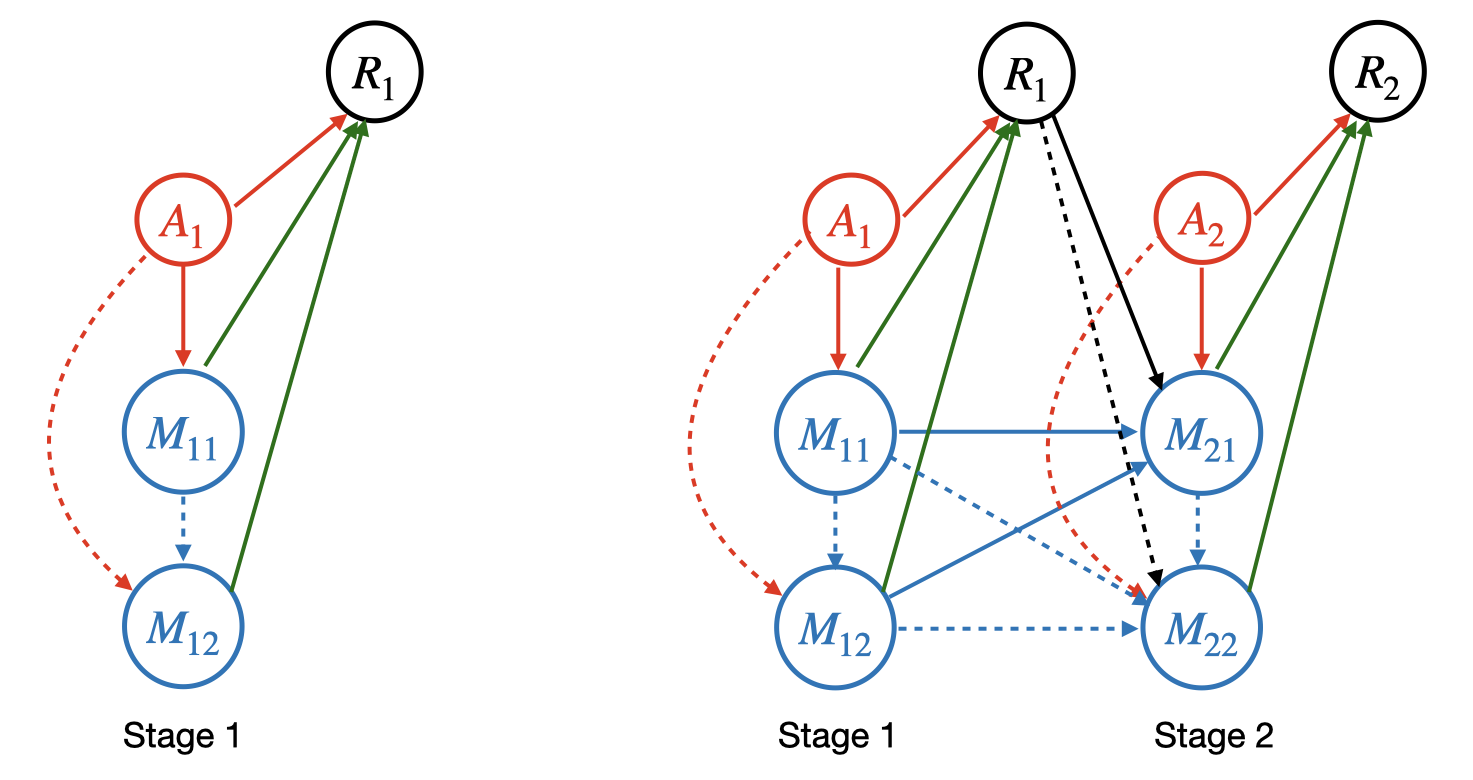}
\caption{An illustrative example with two mediators and two time stages. The left panel shows the DAG in the first stage, and the right panel the first two stages. The dashed lines highlight the paths that go through the intervened mediator, i.e., the second mediator.} 
\label{fig:two-stage}
\end{figure}

To illustrate our theory, we first consider a simple example as shown in Figure \ref{fig:two-stage}, in which there are only $d=2$ mediators and $T=2$ time stages. We then extend our observations to more general cases with $d$ mediators and $T$ stages. 

Suppose we focus on the individual mediation effect for the second mediator, i.e., $j=2$. We begin with the first stage $t=1$, as shown in Figure \ref{fig:two-stage}, left panel. By Definition \ref{def:finite},  
\begin{align} \label{eq:ex-eta12}
\eta_2^{(1)} = \frac{\partial}{\partial a}\mathbb{E}[R_1\mid do(A_1=a)] - \frac{\partial}{\partial a}\mathbb{E}[R_1\mid do(A_1=a, M_{12}=m)] 
\equiv \theta_{A_1\to R_1} - \theta_{A_1\to R_1}^{M_{12}},
\end{align}
where we use the notation $\theta_{A_1\to R_1}$ to denote the total effect from $A_1$ to $R_1$, and use $\theta_{A_1\to R_1}^{M_{12}}$ to denote the total effect from $A_1$ to $R_1$ when the second mediator is intervened in the first stage. 

We compute $\theta_{A_1\to R_1}$ by summing up the effects along all directed paths from $A_1$ to $R_1$, namely, $A_1\to R_1$, $A_1\to M_{11}\to R_1$, $A_1\to M_{12}\to R_1$, and $A_1\to M_{11}\to M_{12}\to R_1$. Meanwhile, we compute $\theta_{A_1\to R_1}^{M_{12}}$ by eliminating all paths that go from $A_1$ to $R_1$ through $M_{12}$. This corresponds to subtracting the effects along $A_1\to M_{12}\to R_1$ and $A_1\to M_{11}\to M_{12}\to R_1$, which leads to
\begin{align} \label{eqn:recursive-A1}
\theta_{A_1\to R_1}^{M_{12}} = \theta_{A_1\to R_1} - \theta_{A_1\to M_{12}}\theta_{M_{12}\to R_1},
\end{align}
where we use the notation $\theta_{A_1\to M_{12}}$ and $\theta_{M_{12}\to R_1}$ to denote the effect from $A_1$ to $M_{12}$, and from $M_{12}$ to $R_1$, respectively. The relation in \eqref{eqn:recursive-A1} has an intuitive interpretation: to evaluate the total effect from $A_1$ to $R_1$ when $M_{12}$ is intervened, we subtract from $\theta_{A_1\to R_1}$ the effects along the paths that go through $M_{12}$. Plugging \eqref{eqn:recursive-A1} into \eqref{eq:ex-eta12}, we obtain that $\eta_2^{(1)}=\theta_{A_1\to M_{12}}\theta_{M_{12}\to R_1}$, which coincides with the classical product type representation of the individual mediation effect in a single-stage analysis \citep{nandy2017joint}.

We next move to the second stage $t=2$, as in Figure~\ref{fig:two-stage}, right panel. By Definition \ref{def:finite},
\begin{align} \label{eq:ex-eta22}
\eta_2^{(2)} & = \eta_2^{(1)} + \bigg\{ \frac{\partial}{\partial a}\mathbb{E}[R_2\mid do(A_1=A_2=a)] \\
& - \frac{\partial}{\partial a}\mathbb{E}[R_2\mid do(A_1=A_2=a, M_{12}=M_{22}=m)] \bigg\} 
\equiv 
\eta_2^{(1)} + \theta_{A_1, A_2\to R_2} - \theta_{A_1,A_2\to R_2}^{M_{12}, M_{22}}
\nonumber
\end{align}
where we use the notation $\theta_{A_1,A_2\to R_2}$ to denote the cumulative total effect of $(A_1,A_2)$ on $R_2$, and use $\theta_{A_1,A_2\to R_2}^{M_{12},M_{22}}$ to denote the cumulative total effect of $(A_1,A_2)$ on $R_2$ when the second mediator is intervened in both the first and second stages. 

Because the treatments are randomly assigned and their parent sets are empty, we have, 
\begin{align} \label{eqn:theta-decomp2}
\theta_{A_1,A_2\to R_2} = \theta_{A_1\to R_2}+\theta_{A_2\to R_2}, \quad\quad
\theta_{A_{1},A_2\to R_2}^{M_{12},M_{22}} = \theta_{A_{1}\to R_2}^{M_{12},M_{22}} + \theta_{A_2\to R_2}^{M_{22}},
\end{align}
where we use the notation $\theta_{A_1\to R_2}$ and $\theta_{A_2\to R_2}$ to denote the total effects from $A_1$ to $R_2$, and from $A_2$ to $R_2$, respectively, use $\theta_{A_2\to R_2}^{M_{22}}$ to denote the effect from $A_2$ to $R_2$ when the second mediator is intervened in the second stage, and use $\theta_{A_1\to R_2}^{M_{12},M_{22}}$ to denote the effect from $A_1$ to $R_2$ when the second mediator is intervened in both stages. 

We compute $\theta_{A_1\to R_2}$ and $\theta_{A_2\to R_2}$ similarly as that for $\theta_{A_1\to R_1}$. We compute $\theta_{A_2\to R_2}^{M_{22}}$ similarly as in \eqref{eqn:recursive-A1}, i.e., $\theta_{A_2\to R_2}^{M_{22}} = \theta_{A_2\to R_2} - \theta_{A_2\to M_{22}}\theta_{M_{22}\to R_2}$. Also, similar to \eqref{eqn:recursive-A1}, we have,
\begin{align} \label{eqn:recursive-A2}
\theta_{A_1\to R_2}^{M_{12},M_{22}} = \theta_{A_{1}\to R_2}^{M_{22}} - \theta_{A_1\to M_{12}} \theta_{M_{12}\to R_2}^{M_{22}},
\end{align}
where we use $\theta_{A_1\to R_2}^{M_{22}}$ and $\theta_{M_{12}\to R_2}^{M_{22}}$ to denote the effects from $A_1$ to $R_2$, and from $M_{12}$ to $R_2$, when the second mediator is intervened in the second stage. Intuitively, $\theta_{M_{12}\to R_2}^{M_{22}}$ can be interpreted as the carryover effect of the second mediator on the outcome from the first stage to the second stage when it is intervened.  We again compute $\theta_{A_1\to R_2}^{M_{22}}$ similarly to \eqref{eqn:recursive-A1}, i.e., $\theta_{A_1\to R_2}^{M_{22}} = \theta_{A_1\to R_2} - \theta_{A_1\to M_{22}}\theta_{M_{22}\to R_2}$. In addition, we compute $\theta_{M_{12}\to R_2}^{M_{22}}$ by eliminating all paths that go from $M_{12}$ to $R_2$ through $M_{22}$, which leads to 
\begin{align} \label{eqn:recursive-M2}
\theta_{M_{12}\to R_2}^{M_{22}} = \theta_{M_{12}\to R_2} - \theta_{M_{12}\to M_{22}} \theta_{M_{22}\to R_2},
\end{align}
where we use $\theta_{M_{12}\to R_2}$, $\theta_{M_{12}\to M_{22}}$, and $\theta_{M_{22}\to R_2}$ to denote the effects from $M_{12}$ to $R_2$, from $M_{12}$ to $M_{22}$, and from $M_{22}$ to $R_2$, respectively. 

Based on the derivations so far, we see that, to evaluate the individual mediation effect, it is crucial to calculate those intermediate quantities, such as $\theta_{A_1\to R_1}, \theta_{A_1\to M_{22}}, \theta_{M_{12}\to M_{22}}$, $\theta_{M_{22}\to R_2}$, among others. Next, we briefly discuss how to evaluate those intermediate quantities after imposing the linear structural equation models \eqref{eq:LSEM} and \eqref{eq:model_setup}. We also discuss some recursive relation that facilitates both the computational and statistical efficiencies. 

First, we note that, under \eqref{eq:LSEM} and \eqref{eq:model_setup}, we can estimate those intermediate quantities through linear regressions. For instance, we can estimate $\theta_{A_1\to R_1}$ as the coefficient of $A_1$ by linearly regressing $R_1$ onto $A_1$. Meanwhile, we can estimate $\theta_{M_{22}\to R_2}$ as the coefficient of $M_{22}$ by linearly regressing $R_2$ onto $M_{22}$, however, with some additional covariate adjustment. This is because, unless the two mediators $(M_{21},M_{22})$ are conditionally independent given $(M_{11}, M_{12}, R_1,A_2)$, we need to adjust for a set of covariates that satisfy Pearl's backdoor criterion \citep{pearl2000do}. In other words, we should block the effects flowing from $(A_2, M_{11}, M_{12}, M_{21}, R_1)$ to $M_{22}$. Therefore, we need to adjust for the covariate set $\Pa(M_{22}) \cup {M_{11}}\cup M_{12}\cup R_1 \cup A_2$ in this regression. 

Second, we note that, the set of intermediate quantities involve both within-stage quantities such as $\theta_{A_1\to R_1}, \theta_{M_{22}\to R_2}$, as well as cross-stage quantities such as $\theta_{A_1\to M_{22}}, \theta_{M_{12}\to M_{22}}$. There is some useful recursive relation between the within-stage and cross-stage quantities under the linear structural equation models \eqref{eq:LSEM} and \eqref{eq:model_setup}. For instance, 
\begin{align} \label{eqn:transition-a1m2}
\btheta_{A_1\to \bM_{2}} = \bGamma_{12}\btheta_{A_1\to \bM_1}+\bzeta_{12}\theta_{A_1\to R_1}, 
\end{align}
where $\btheta_{A_1\to \bM_2}=(\theta_{A_1\to M_{21}},\theta_{A_1\to M_{22}})\trans$, and $\btheta_{A_1\to \bM_1}=(\theta_{A_1\to M_{11}},\theta_{A_1\to M_{12}})\trans$. This suggests that {the cross-stage carryover effect} from $A_1$ to $\bM_2 = (M_{21}, M_{22})\trans$ is a combination of its within-stage effect on $\bM_1 = (M_{11}, M_{12})\trans$ and on $R_1$, respectively, whereas the coefficients $(\bGamma_{12}, \bzeta_{12})$ can be viewed as the weights for such a cross-stage transition. In our estimation algorithm, we first estimate the within-stage quantities via linear regressions, then update the cross-stage quantities following \eqref{eqn:transition-a1m2}. As we show later, the relation such as \eqref{eqn:transition-a1m2} not only expedites the computation, but also improves the estimation efficiency by leveraging more information from the conditional model \eqref{eq:model_setup}.

In summary, this simple example reveals a number of important relations. First, we see that those intermediate quantities form the building blocks for our evaluation of the individual mediation effect. Second, \eqref{eqn:recursive-A1}, \eqref{eqn:recursive-A2}, and \eqref{eqn:recursive-M2} suggest some useful recursive representations that in effect reduce the number of mediators in the superscript that are intervened upon. Third, under models \eqref{eq:LSEM} and \eqref{eq:model_setup}, we can estimate those intermediate quantities through linear regressions, with possibly backdoor adjustment. Finally, we can further improve both the computational and statistical efficiencies through another set of recursive relations between the within-stage and cross-stage quantities. All these observations are crucial for our derivation of the expression for the individual mediation effect.

\subsection{Intermediate quantities}
\label{ssec:intermediate}

We now formally define the set of all intermediate quantities that are needed for the evaluation of the individual mediation effect. We then discuss how to estimate those quantities. 

For any $1 \leq s \leq t \leq T$, and $j = 1, \ldots, d$, define
\vspace{-0.01in}
\begin{align} \label{eq:intermediate}
\begin{split}
& \theta_{A_s\to R_t} \equiv \frac{\partial}{\partial a}\Ebb[R_t\mid do(A_s=a)] \in \Rbb, \quad\,\
\btheta_{M_{sj}\to \bM_t} \equiv \frac{\partial}{\partial m}\Ebb[\bM_t\mid do(M_{sj}=m)] \in \Rbb^{d}, \\
& \btheta_{A_s\to \bM_t} \equiv \frac{\partial}{\partial a}\Ebb[\bM_t\mid do(A_s=a)] \in \Rbb^{d}, \;
\theta_{M_{sj}\to R_t} \equiv \frac{\partial}{\partial m}\Ebb[R_t\mid do(M_{sj}=m)] \in \Rbb. 
\end{split}
\end{align}

These intermediate quantities in \eqref{eq:intermediate} can be obtained through linear regressions, either directly, or by some backdoor covariate adjustment. In particular, $\theta_{A_s\to R_t}$ can be obtained as the coefficient of $A_s$ by linearly regressing $R_t$ onto $A_s$ with an intercept, and we write this coefficient as $\beta_{A_s,R_t}$. Similarly, $\btheta_{A_s\to \bM_t}$ can be obtained as the coefficient of $A_s$ by linearly regressing $\bM_t$ onto $A_s$, denoted as $\bbeta_{A_s,\bM_t}$. Meanwhile, $\theta_{M_{sj}\to R_t}$ can be obtained as the coefficient of $M_{sj}$ by linearly regressing $R_t$ onto $M_{sj}$, along with the adjusted covariate set, $\Pa(M_{sj})\cup {\bM_{s-1} }\cup R_{s-1}\cup A_s$. Similarly, $\theta_{M_{sj}\to \bM_t}$ can be obtained as the coefficient of $M_{sj}$ by linearly regressing $\bM_t$ onto $M_{sj}$, along with the adjusted covariate set, $\Pa(M_{sj})\cup {\bM_{s-1} }\cup R_{s-1}\cup A_s$. Putting together, we have, 
\begin{align} \label{eq:within-stage}
\begin{split}
& \theta_{A_s\to R_t} = \beta_{A_s,R_t}, \quad\quad 
\btheta_{M_{sj}\to \bM_t} = \bbeta_{M_{sj}, \bM_t\mid \Pa(M_{sj})\cup {\bM_{s-1} }\cup R_{s-1}\cup A_s}, \\
& \btheta_{A_s\to \bM_t} = \bbeta_{A_s, \bM_t},  \quad\,
\theta_{M_{sj}\to R_t} = \beta_{M_{sj}, R_t\mid \Pa(M_{sj})\cup \bM_{s-1}\cup R_{s-1}\cup A_s}. 
\end{split}
\end{align}

Moreover, similar to \eqref{eqn:transition-a1m2}, we obtain the following recursive relations between the within-stage quantities and the cross-stage quantities. That is, under models \eqref{eq:LSEM} and \eqref{eq:model_setup}, 
\begin{equation}\label{eq:recursion_total_effects}
\begin{split}
& \theta_{A_s\to R_{t}} = \zeta_{2t} \theta_{A_s\to R_{t-1}}+\bkappa_t\trans \btheta_{A_s\to \bM_{t}}+\bgamma_{2t}\trans \btheta_{A_s \to \bM_{t-1}} , \\
& \btheta_{A_s\to \bM_{t}} = \bGamma_{1t} \btheta_{A_s\to \bM_{t-1}}  + \bzeta_{1t} \theta_{A_s\to R_{t-1}}, \\
& \theta_{M_{sj}\to R_t} = \zeta_{2t} \theta_{M_{s j}\to R_{t-1}} + \bkappa_t\trans \btheta_{M_{sj}\to \bM_{t}} + \bgamma_{2t}\trans\btheta_{M_{sj}\to \bM_{t-1}}, \\
& \btheta_{M_{sj}\to \bM_{t}} = \bGamma_{1t} \btheta_{M_{sj}\to \bM_{t-1}}+ \bzeta_{1t} \theta_{M_{sj}\to R_{t-1}}.
\end{split}
\end{equation}

In our implementation, we first estimate the within-stage quantities, $\{ \theta_{A_t \to R_t}, \theta_{M_{tj} \to R_t}$, $\btheta_{A_t \to \bM_t}, \btheta_{M_{tj}\to \bM_t} \}$, using \eqref{eq:within-stage}, then estimate the cross-stage quantities, $\{ \theta_{A_s\to R_t}, \theta_{M_{sj}\to R_t}$, $\btheta_{A_s\to R_t}, \btheta_{M_{sj}\to \bM_t} \}$, for $s=1, \ldots, t-1$, using \eqref{eq:recursion_total_effects}. This helps to improve both the estimation efficiency and the computation efficiency.

\subsection{Individual mediation effect in a finite-horizon setting}
\label{ssec:estinf-finite}

We next derive the expression of the individual mediation effect through the intermediate quantities in \eqref{eq:intermediate} under the finite-horizon setting. 

We first sketch the key ideas of the derivation here, and relegate more details to the Supplementary Appendix. By Definition~\ref{def:finite}, the individual mediation effect of the $j$th mediator across $t$ stages, $j = 1, \ldots, d$, $t = 1, \ldots, T$, can be expressed as
\begin{equation}\label{eq:finite_eta_expression}
\eta_j^{(t)} = \eta_j^{(t-1)} + \theta_{A_1,\dots,A_t\to R_t} - \theta_{A_1,\dots,A_t\to R_t}^{M_{1j},\dots, M_{tj}}. 
\end{equation}

Because all the treatments $(A_1,\ldots,A_t)$ are randomly assigned, and thus are independent of each other and all other covariates, similar to \eqref{eqn:theta-decomp2}, we have,
\begin{align} \label{eqn:theta-decomp}
\begin{split}
\theta_{A_1,\dots,A_t\to R_t} &= \theta_{A_1\to R_t}+\theta_{A_2\to R_t} + \dots + \theta_{A_t\to R_t}, \\
\theta_{A_1,\dots,A_t\to R_t}^{M_{1j},\dots,M_{tj}} &= \theta_{A_1\to R_t}^{M_{1j},\dots,M_{tj}} + \theta_{A_2\to R_t}^{M_{2j},\dots,M_{tj}} + \dots + \theta_{A_t\to R_t}^{M_{tj}}.
\end{split}
\end{align}

Then, similar to \eqref{eqn:recursive-A1}, \eqref{eqn:recursive-A2}, and \eqref{eqn:recursive-M2}, we obtain the following recursive relations that help reduce the number of mediators being intervened upon. That is, for any $s = 1, \ldots, t-1$,
\begin{align} 
\theta_{A_s\to R_t}^{M_{sj},\ldots,M_{tj}} & 
= \theta_{A_s\to R_t}^{M_{(s+1)j},\ldots,M_{tj}} - \theta_{A_s\to M_{sj}}\theta_{M_{sj}\to R_t}^{M_{sj},\ldots, M_{tj}} = \cdots \nonumber \\
& = \theta_{A_s\to R_t}  - \sum_{i=s}^{t}\theta_{A_s\to M_{ij}}\theta_{M_{ij}\to R_t}^{M_{ij},\ldots,M_{tj}};  \label{eqn:recursive-A} \\
\theta_{M_{s j}\to R_t}^{M_{sj},\ldots, M_{tj}} & 
= \theta_{M_{sj}\to R_t}^{M_{(s+2)j},\ldots,M_{tj}} - \theta_{M_{sj}\to M_{(s+1)j}}\theta_{M_{(s+1)j}\to R_t}^{M_{(s+1)j},\ldots, M_{tj}} = \cdots \nonumber \\
& = \theta_{M_{sj}\to R_t} - \sum_{i=s+1}^{t} \theta_{M_{sj}\to M_{ij}} \theta_{M_{ij}\to R_t}^{M_{ij},\ldots, M_{tj}}. \label{eqn:recursive-M}  
\end{align}

Combining \eqref{eq:finite_eta_expression}, \eqref{eqn:theta-decomp}, \eqref{eqn:recursive-A}, and \eqref{eqn:recursive-M}, we obtain the following theorem with respect to the identification of individual mediation effect in finite-horizon settings. We relegate a more detailed derivation to the Supplementary Appendix.

\begin{theorem}[Individual mediation effect for finite-horizon]\label{thm:eta_finite}
Suppose $\{A_t\}_{t \geq 1}$ are randomly assigned treatments and satisfy \eqref{eq:rdmtrt}, and $\{ \bM_t,R_t \}_{t \geq 1}$ follow \eqref{eq:LSEM} and \eqref{eq:model_setup}. Then, 
\begin{equation}\label{eq:effects_finite_horizon}
\begin{split}
\eta_j^{(t)}
& = \eta_j^{(t-1)} + \sum_{s=1}^t\sum_{i=s}^t\theta_{A_s\to M_{ij}} \theta_{M_{ij}\to R_t}^{M_{ij},\dots,M_{tj}}, 
\; {t=1,\ldots,T}, 
\end{split}
\end{equation}
where $\theta_{M_{s j}\to R_t}^{M_{sj},\dots, M_{tj}}$ is computed following \eqref{eqn:recursive-M}, and we set $\eta_j^{(0)}=0$. 
\end{theorem}

As for estimation, based on Theorem \ref{thm:eta_finite}, we estimate the individual mediation effect in a recursive manner. That is, for stage $t-1$, we first estimate the weight matrix $\bW_t$ in \eqref{eq:LSEM}, and the parameters $\bTheta_{1t},\bTheta_{2t}$ in \eqref{eq:model_setup} for stage $t$. Estimation of $\bW_t$ is needed for determining the parent set of each mediator, which in turn is used for backdoor covariate adjustment. {There are multiple algorithms available to estimate $\bW_t$, including those developed in single-time-point settings \citep{zheng2018dags, yuan2019constrained, bello2022dagma}, and those developed in time-varying settings \citep{hyvarinen2010estimation,malinsky2018causal,pamfil2020dynotears}. We employ the DAGMA algorithm recently proposed by \cite{bello2022dagma} for its simplicity and effectiveness. Recall that, in the finite-horizon setting, we allow the DAG to be nonstationary over time. We thus estimate the time-specific DAGs using DAGMA, with data pooled over multiple subjects but restricted to that time point only.} We then estimate the within-stage quantities $\{\theta_{A_t\to R_t}, \theta_{M_{tj} \to R_t}$, $\btheta_{A_t \to \bM_t}, \btheta_{M_{tj}\to \bM_t} \}$ using \eqref{eq:within-stage}, and estimate the cross-stage quantities, $\{\theta_{A_s\to R_t}, \theta_{M_{sj}\to R_t}$, $\btheta_{A_{s}\to \bM_t}, \btheta_{M_{sj}\to \bM_t} \}$, for $s=1, \ldots, t-1$, using \eqref{eq:recursion_total_effects}. Finally, we plug-in these estimators into  \eqref{eq:effects_finite_horizon} to estimate the individual mediation effect. 

\begin{algorithm}[t!]
\caption{Estimation and inference for the finite-horizon setting.}
\label{alg:finite}
\begin{algorithmic}[1]
\Require Observed data $\{(A_{it},\bM_{it},R_{it}): i = 1, \ldots, n, t = 1, \ldots, T\}$, {the number of bootstrap samples $B$, and the significance level $\alpha$}.
\Ensure Individual mediation effect $\widehat{\eta}^{(t)}_j$, for $t = 1, \ldots, T, j = 1, \ldots, d$, and their CIs. 
\State Estimate within-stage quantities $(\widehat\theta_{A_1\to R_1}, \widehat\theta_{M_{1j}\to R_1}, \widehat\btheta_{A_1\to\bM_1}, \widehat\btheta_{M_{1j}\to\bM_1})$, and compute $\widehat{\eta}_j^{(1)} = \widehat\theta_{A_1\to M_{1j}}\widehat\theta_{M_{1j}\to  R_1}$.
\For{$t = 2,\dots,T$}
\State Estimate parameters $\widehat\bW_t, \widehat\bTheta_{1t}, \widehat\bTheta_{2t}$ in models \eqref{eq:LSEM} and \eqref{eq:model_setup}.
\State Estimate within-stage quantities $(\widehat\theta_{A_{t}\to R_t}, \widehat{\theta}_{M_{tj} \to R_t}, \widehat{\btheta}_{A_t \to \bM_t}, \widehat{\btheta}_{M_{tj}\to\bM_t})$, 
\For{$s = 1, \ldots, t-1$}
\State Estimate cross-stage quantities $(\widehat\theta_{A_{s}\to R_t}, \widehat\theta_{M_{sj}\to R_t}, \widehat\btheta_{A_s\to \bM_t}, \widehat\btheta_{M_{sj}\to \bM_t})$ using \eqref{eq:recursion_total_effects}.
\EndFor
\State Compute $\widehat\eta_j^{(t)}$ using \eqref{eq:effects_finite_horizon}.
\EndFor
{\For{$b=1,\cdots,B$}
\State Sample $n$ trajectories from the observed data with replacement.
\State Repeat Lines 1 to 9 to compute $\widehat\eta_j^{(t,b)}$ using the bootstrap samples.
\EndFor
\For{$t = 1,\dots,T$}
\State Construct the CI $\left[ \widehat\eta_j^{(t,L)},\widehat\eta_j^{(t,U)} \right]$ for $\eta_j^{(t)}$, where $\widehat\eta_j^{(t,L)}$ and $\widehat\eta_j^{(t,U)}$ are the empirical lower and upper 
\State $\alpha/2$ quantiles of $\{\widehat\eta_j^{(t,b)}\}_{b=1}^B$, respectively. 
\EndFor
}
\end{algorithmic}
\end{algorithm}

{As for inference, based on Theorem \ref{theory:finite} in Section \ref{sec:theory}, our plug-in estimator is asymptotically normal. This motivates us to employ the nonparametric bootstrap approach \citep{efron1979bootstrap} to conduct statistical inference. Specifically, we first sample $n$ trajectories from the observed data with replacement. We next refit the weight matrix $\bW_t$ and re-estimate the within-stage and cross-stage quantities using the bootstrap samples. During the refitting, there are two options. One is to use the bootstrap samples to refit the coefficients of the nonzero entries in the weight matrix only, without re-estimating the structure of $\bW_t$. The other is to refit both the structure of $\bW_t$ and its nonzero entries. When the structure learning algorithm is consistent, the two solutions are asymptotically equivalent. For finite samples, the former is computationally more efficient, while the latter is expected to be more accurate. In our implementation, we adopt the first option due to its simplicity, and our numerical experiments suggest that it works well empirically. We next construct the plug-in estimators using these refitted parameters. Finally, we use the empirical lower and upper $\alpha/2$ quantiles of the plug-in estimators to construct the confidence interval under a given significance level $\alpha$.} 

Algorithm~\ref{alg:finite} summarizes our estimation and inference procedure.

\subsection{Individual mediation effect in an infinite-horizon setting}
\label{ssec:estinf-infinite}

We next derive the expression of the individual mediation effect under the infinite-horizon setting. Unlike the finite-horizon setting, we now require the Markov mediation process $\{A_t,\bM_t, R_t\}_{t\geq 1}$ to be stationary. This is to ensure the existence of the limit in Definition \ref{def:infinite}. Correspondingly, the parameters $\bW, \bTheta_{1},\bTheta_{2}$ in \eqref{eq:LSEM} and \eqref{eq:model_setup} remain the same across different time stages, which leads to the following relations. For any $1\leq s\leq t\leq T$, and $j = 1,\ldots, d$, 
\begin{align*} 
\begin{split}
& \theta_{A_s \to R_{t}} = \theta_{A_{s+1} \to R_{t+1}} = \ldots = \theta_{A_{s+T-t} \to R_{T}}, \; \btheta_{M_{sj} \to \bM_{t}} = \btheta_{M_{(s+1)j} \to \bM_{t+1}} = \ldots = \btheta_{M_{(s+T-t)j} \to \bM_{T}}, \\
& \btheta_{A_s \to \bM_{t}} = \btheta_{A_{s+1} \to \bM_{t+1}} = \ldots = \btheta_{A_{s+T-t} \to \bM_{T}}, \;
\theta_{M_{sj} \to R_{t}} = \theta_{M_{(s+1)j} \to R_{t+1}} = \ldots = \theta_{M_{(s+T-t)j} \to R_{T}}.
\end{split}
\end{align*}
Based on this observation, we obtain a simplified representation for $\eta_j^{(T)}$ as,
\begin{align*} 
\eta_j^{(T)} = \theta_{M_{1j}\to R_{1}} \sum_{t=1}^T(T-t+1) \theta_{A_{1}\to M_{tj}}+ \sum_{t=1}^{T-1}\theta_{M_{1j}\to R_{t+1}}^{M_{1j},\dots,M_{(t+1)j}}\left(\sum_{s=1}^{T-t} (T-t-s + 1) \theta_{A_{1}\to M_{sj}} \right). 
\end{align*}

By dividing the right-hand-side by $T$ and taking the limit $T \to \infty$, we obtain the following theorem for identifying the individual mediation effect in infinite-horizon settings. We relegate a more detailed derivation to the Supplementary Appendix.

\begin{theorem}[Individual mediation effect for infinite-horizon]\label{thm:eta_infinite}
Suppose $\{A_t\}_{t \geq 1}$ are randomly assigned treatments and satisfy \eqref{eq:rdmtrt}, $\{ \bM_t,R_t \}_{t \geq 1}$ follow \eqref{eq:LSEM} and \eqref{eq:model_setup}, the Markov process $\{A_t,\bM_t,R_t\}_{t\geq 1}$ is stationary, and the parameters $\{\bW, \bTheta_{1},\bTheta_{2}\}$ in \eqref{eq:LSEM} and \eqref{eq:model_setup} are time-invariant. Then, 
\begin{equation}\label{eq:effects_infinite_horizon}
\begin{split}
\eta_j^{(\infty)} \equiv \underset{T\to \infty}{\lim }\frac{\eta_j^{(T)}}{T}
= \theta_{M_{1j}\to R_1} B_{1j} + (1 + B_{2j})^{-1}(B_{5j} - \theta_{M_{1j}\to R_1} B_{4j}),
\end{split}
\end{equation}
where $B_{1j}, B_{2j}, B_{4j}$ are the $j$th element of $\bB_1, \bB_2, \bB_4$, respectively,
\begin{equation}\label{eq:Bs} 
\begin{split}
& \;\;\;\, \bB_1 = \left\{(1-\zeta_2)(\bI-\bGamma_{1})-\bzeta_1(\bkappa+\bgamma_{2})\trans \right\}^{-1} \left\{(1-\zeta_2)\bdelta_1 + \delta_2\bzeta_1 \right\} \in \Rbb^{d}, \\
& \begin{pmatrix}
\bB_2 \\ B_{3j}
\end{pmatrix} = (\bI - \bB_6 - \bB_7)^{-1} \bB_7 
\begin{pmatrix} \btheta_{M_{1j}\to \bM_1} \\ \theta_{M_{1j}\to R_1} \end{pmatrix} \in \Rbb^{d+1}, \\
& \begin{pmatrix}
\bB_4 \\ B_{5j}
\end{pmatrix} = (\bI - \bB_6 - \bB_7)^{-1} \bB_7 
\begin{pmatrix}\btheta_{M_{1j}\to \bM_1} \\ \theta_{M_{1j}\to R_1} \end{pmatrix} B_{1j} \in \Rbb^{d+1},
\end{split}
\end{equation}
$\bB_6 = \begin{pmatrix} \bzero_{d\times d} & \bzero_{d} \\ \bkappa\trans & 0 \end{pmatrix}\in\Rbb^{(d+1)\times (d+1)}$, and $\bB_7 = \begin{pmatrix} \bGamma_1 & \bzeta_1 \\ \bgamma_2\trans & \zeta_2 \end{pmatrix}\in\Rbb^{(d+1)\times (d+1)}$. 
\end{theorem}

As for estimation, based on Theorem \ref{thm:eta_infinite}, we pool the data across all $T$ stages to estimate the model parameters $\bW, \bTheta_1, \bTheta_2$. {Recall that, in the infinite-horizon setting, we require the DAG to be stationary over time. We thus apply DAGMA to the data pooled across all time points to learn this time-invariant DAG. This is different from the finite-horizon setting where the model parameters can differ from one stage to another, while in the infinite-horizon setting, they remain the same.} We then estimate the quantities $\bB_1$ to $\bB_7$ in \eqref{eq:Bs} by plugging in the estimates of the corresponding model parameters, and we estimate the individual mediation effect using \eqref{eq:effects_infinite_horizon}. {As for inference, we adopt a similar nonparametric bootstrap procedure as in the finite-horizon setting.} Algorithm~\ref{alg:infinite} summarizes our estimation and inference procedure.

\begin{algorithm}[t!]
\caption{Estimation for the infinite-horizon setting.}
\label{alg:infinite}
\begin{algorithmic}[1]
\Require Observed data $\{(A_{it},\bM_{it},R_{it}): i = 1, \ldots, n, t = 1, \ldots, T\}$, {the number of bootstrap samples $B$, and the significance level $\alpha$}.
\Ensure Estimated individual mediation effect $\widehat{\eta}_j^{(\infty)}$, for $j = 1, \ldots, d$. 
\State Pool data across all $t$ stages, $t = 1, \ldots, T$.
\State Estimate parameters $\widehat{\bW}, \widehat\bTheta_1, \widehat\bTheta_2$ using the pooled data.
\State Estimate $\btheta_{M_{1j}\to\bM_1}$ and $\theta_{M_{1j}\to R_1}$ using the pooled data.
\State Compute $\bB_1$ to $\bB_7$ using \eqref{eq:Bs}.
\State Compute $\widehat{\eta}_j^{(\infty)}$ using \eqref{eq:effects_infinite_horizon}.
{\For{$b=1,\cdots,B$}
\State Sample $n$ trajectories from the observed data with replacement.
\State Repeat Lines 1 to 5 to compute $\widehat\eta_j^{(\infty,b)}$ using the bootstrap samples.
\EndFor
\State Construct the CI $\left[ \widehat\eta_j^{(\infty,L)},\widehat\eta_j^{(\infty,U)} \right]$ for $\eta_j^{(\infty)}$, where $\widehat\eta_j^{(\infty,L)}$ and $\widehat\eta_j^{(\infty,U)}$ are the empirical lower and upper $\alpha/2$ quantiles of $\{\widehat\eta_j^{(\infty,b)}\}_{b=1}^B$, respectively.}
\end{algorithmic}
\end{algorithm}

\subsection{Asymptotic theory}
\label{sec:theory}

We establish the asymptotic properties of our estimator of the individual mediation effect. We first present a set of regularity conditions. 

\begin{assumption}[Invertibility]\label{asp:invertibility}
	{Let $\bX_{t}\equiv (1, A_{t}, \bM_{t-1}\trans, R_{t-1},\bM_t\trans)\trans\in\mathbb{R}^{2d+3}$. There exists some constant $C>0$ such that $\lambda_{\min}(\mathbb{E} \bX_t \bX_t^\top)\ge C$ for any $t$, where $\lambda_{\min}(\bullet)$ denotes the minimum eigenvalue of a given matrix.}
\end{assumption}

\begin{assumption}[Error residuals]\label{asp:finite_fourth_m}
{(i) The error terms $\epsilon_{w1t}$, $\epsilon_{w2t}$, $\cdots$, $\epsilon_{wdt}$ in \eqref{eq:LSEM} are  jointly normally distributed and independent, for $t=1, \ldots, T$. In addition, their variances are constant, in that $\mbox{Var}(\epsilon_{w1t})=\mbox{Var}(\epsilon_{w2t})=\ldots=\mbox{Var}(\epsilon_{wdt})$.} (ii) The error terms $\epsilon_{rt}$ in \eqref{eq:model_setup} satisfy that $\Ebb(\epsilon_{rt}^4) < \infty$, for $t=1, \ldots, T$. 
\end{assumption}

\begin{assumption}[Structure learning consistency]\label{asp:DAG_consistency}
The estimated DAG is a consistent estimator of the true underlying DAG; i.e., (i) for the finite-horizon setting, for $t=1,\dots,T$, $\Pbb(\widehat{\mathcal{G}}_t \neq \mathcal{G}_{t}) \to 0$ as $n \to \infty$, where $\mathcal{G}_{t}$ is the true DAG in stage $t$; (ii) for the infinite-horizon setting, $\Pbb(\widehat{\mathcal{G}} \neq \mathcal{G}) \to 0$ as $T \to \infty$,  where $\mathcal{G}$ is the true time-invariant DAG. 
\end{assumption}

\begin{assumption}[Stationarity]\label{asp:stationary}
For the infinite-horizon setting, the process $\{A_t,\bM_t,R_t\}_{t\geq 1}$ is a strictly stationary $\beta$-mixing process.
\end{assumption}

We give some remarks about these conditions. {Assumption \ref{asp:invertibility} guarantees the identifiability of model parameters in~\eqref{eq:model_setup}.} {Assumption \ref{asp:finite_fourth_m}(i) is commonly imposed for identifying DAG structures; see, e.g.,  \citet{peters2014identifiability,nandy2017joint,yuan2019constrained,shi2022testing}.} Assumption \ref{asp:finite_fourth_m}(ii) requires the fourth moments of the errors to be finite, which is mild. Assumption \ref{asp:DAG_consistency} holds for numerous DAG estimation algorithms, including the one by \cite{bello2022dagma} that we use in our implementation. Assumption~\ref{asp:stationary} is required for the infinite-horizon setting only, and the $\beta$-mixing condition is again commonly imposed; see, e.g., \citet{Bradley2005}. Overall, these regularity conditions are reasonable. 

We next establish the asymptotic normality of our individual mediation effect estimator for both the finite-horizon and infinite-horizon settings. 
  
\begin{theorem}[Asymptotic distribution for finite-horizon]\label{theory:finite}
Suppose Assumptions \ref{asp:finite_fourth_m} and \ref{asp:DAG_consistency}(i) hold. Then, for $t=1,\ldots,T$, 
\vspace{-0.01in}
\begin{equation*}
\sqrt{n}(\widehat{\eta}_j^{(t)} - \eta_{j}^{(t)}) \overset{d}{\to}\mathcal{N}(0, \sigma_{tj}^2), \ \text{as} \ n \to \infty, 
\vspace{-0.01in}
\end{equation*}
where $\sigma_{tj}^2$ denotes the asymptotic variance of $\widehat{\eta}_j^{(t)}$. Its form is given in the Supplementary Material, and it can be consistently estimated via the bootstrap method.
\end{theorem}

\begin{theorem}[Asymptotic distribution for infinite-horizon]\label{theory:infinite}
Suppose Assumptions \ref{asp:finite_fourth_m},  \ref{asp:DAG_consistency}(ii) and \ref{asp:stationary} hold. Then,
\vspace{-0.05in}
\begin{equation*}
\sqrt{{n}T}({\widehat{\eta}^{(\infty)}_j - \eta^{(\infty)}_{j}}) \overset{d}{\to}\mathcal{N}(0, \sigma_j^2), \ \text{as} \ {nT \to \infty},
\vspace{-0.05in}
\end{equation*}
{where $\sigma_j^2$ denotes the asymptotic variance of $\widehat{\eta}_j^{(\infty)}$. Its definition is given in the Supplementary Material, and it can be consistently estimated using the bootstrap method.}
\end{theorem}

We briefly remark that Theorem \ref{theory:finite} requires the number of realizations $n$ of $(A_{t},\bM_{t},R_{t})$ to diverge to infinity for every $t=1,\ldots,T$, with $T$ being finite. Meanwhile, Theorem \ref{theory:infinite} requires either $n$ or the number of time points $T$ or both to diverge to infinity.

\section{Simulations}
\label{sec:simulation}

In this section, we investigate the empirical performance of our proposed method through intensive simulations. We also compare with some baseline methods.

\subsection{Simulation setup}
\label{ssec:sim_setup}

We generate $n$ copies of random samples $\{A_t,\bM_t,R_t\}_{t=1}^T$ following models \eqref{eq:LSEM} and \eqref{eq:model_setup}, i.e., 
\begin{align*}
\bM_t &=(\bI - \bW\trans)^{-1}\bepsilon_{wt} + \balpha_{1t} +\bdelta_{1t}A_t+\bGamma_{1t}\bM_{t-1}+\bm{\zeta}_{1t} R_{t-1}, \\
R_t &= \alpha_{2t} + \delta_{2t} A_t + \bgamma_{2t}\trans \bM_{t-1} + \zeta_{2t} R_{t-1} +  \bkappa_t\trans \bM_{t} + \epsilon_{rt}. 
\end{align*}
We generate the sequence of treatments $A_t \overset{iid}{\sim}\text{Bernoulli}(0.5)$, $t=1,\dots,T$, and the error terms $\bepsilon_{wt}$ and $\epsilon_{rt}$ from a standard normal distribution. Let $\bTheta_{1t}=(\balpha_{1t},\bdelta_{1t},\bGamma_{1t},\bzeta_{1t})\trans\in\Rbb^{(d+3)\times d}$ and $\bTheta_{2t}=(\alpha_{2t},\delta_{2t},\zeta_{2t},\bgamma_{2t}\trans,\bkappa_t\trans)\trans \in \Rbb^{(2d+3)}$ collect the model parameters, and we generate the entries of $\bTheta_{1t}, \bTheta_{2t}$ from a uniform distribution on $(-0.5,0.5)$. For the finite-horizon setting, we generate different $\bTheta_{1t},\bTheta_{2t}$ for different time points $t$, whereas for the infinite-horizon setting, we only generate one copy of $\bTheta_{1t},\bTheta_{2t}$, and keep them fixed across all time points. We fix $d=3$, and generate the matrix $\bW \in \Rbb^{3 \times 3}$ in  two steps. We first begin with a zero matrix, then replace every entry $(\bW)_{ij},i<j$, by the product of two random variables $E_{ij}^{(1)} \times E_{ij}^{(2)}$, where $E_{ij}^{(1)}$ is a Bernoulli variable with probability 0.9, indicating a random directed edge is added from mediator $M_{i}$ to $M_{j}$, and $E_{ij}^{(2)}$ is the edge weight, which is randomly drawn from a uniform distribution on $[-0.9,-0.5]\cup[0.5,0.9]$. Following this generation process, we obtain $\bW=\begin{pmatrix}
0 &-0.80 &0.61 \\
0 &0 &-0.82 \\
0 &0 &0
\end{pmatrix}$. We set the initial values of mediators and outcome all equal to $0$. 

We recognize that there is no existing solution in the literature for this problem. Instead, we compare with two baseline solutions. One method is termed ``independent time points'', which ignores all temporal dependence across different time points. That is, it estimates the individual mediation effect at every single time stage with stage-specific data, without taking into account the dependence to prior stages nor the carryover effects. The other method is termed ``independent mediators'', which ignores all dependence among the multivariate mediators, and essentially treats $\bW$ as a zero matrix. We evaluate all estimation methods using three criteria: the estimation bias, the empirical standard error (SE), {and the coverage probability (CP). For inference, we set the number of bootstrapped samples to be $500$ and the significance level $\alpha$ to be $0.05$.} We employ DAGMA with linear models for structural learning, and the corresponding hyperparameters, \texttt{w\_threshold} and \texttt{lambda1}, are set to $0.1$ and $0.0$, respectively.

\subsection{Finite-horizon setting}
\label{ssec:sim_finite}

\begin{table}[t!]
\centering
\caption{{Simulations for the finite-horizon setting: the bias, the empirical standard error (SE) and the coverage probability (CP) between the estimated and true individual mediator effects with varying number of time points $T$ and number of subjects $n$.}}
\label{table:finite_sim}
\begin{tabular}{ccc|rrr|rrr|rrr}
\hline
\multicolumn{3}{c|}{n}                                                                                                                         & \multicolumn{3}{c|}{100}                                                     & \multicolumn{3}{c|}{250}                                                     & \multicolumn{3}{c}{500}                                                     \\
Method                                                                                                  & T                   & Param          & \multicolumn{1}{c}{Bias} & \multicolumn{1}{c}{SE} & \multicolumn{1}{c|}{CP} & \multicolumn{1}{c}{Bias} & \multicolumn{1}{c}{SE} & \multicolumn{1}{c|}{CP} & \multicolumn{1}{c}{Bias} & \multicolumn{1}{c}{SE} & \multicolumn{1}{c}{CP} \\ \hline
\multicolumn{1}{c|}{\multirow{9}{*}{\begin{tabular}[c]{@{}c@{}}Proposed\\ method\end{tabular}}}         & \multirow{3}{*}{10} & $\eta_1^{(T)}$ & 0.000                    & 0.594                   & 0.972                   & 0.003                    & 0.351                   & 0.966                   & 0.008                    & 0.249                   & 0.960                  \\
\multicolumn{1}{c|}{}                                                                                   &                     & $\eta_2^{(T)}$ & -0.016                   & 0.533                   & 0.960                   & -0.036                   & 0.283                   & 0.970                   & -0.006                   & 0.209                   & 0.952                  \\
\multicolumn{1}{c|}{}                                                                                   &                     & $\eta_3^{(T)}$ & 0.007                    & 0.696                   & 0.964                   & 0.023                    & 0.420                   & 0.962                   & 0.011                    & 0.312                   & 0.950                  \\ \cline{2-12} 
\multicolumn{1}{c|}{}                                                                                   & \multirow{3}{*}{20} & $\eta_1^{(T)}$ & -0.046                   & 1.960                   & 0.950                   & 0.008                    & 1.215                   & 0.946                   & -0.023                   & 0.840                   & 0.950                  \\
\multicolumn{1}{c|}{}                                                                                   &                     & $\eta_2^{(T)}$ & -0.037                   & 1.134                   & 0.970                   & -0.018                   & 0.717                   & 0.944                   & -0.021                   & 0.490                   & 0.948                  \\
\multicolumn{1}{c|}{}                                                                                   &                     & $\eta_3^{(T)}$ & 0.027                    & 1.236                   & 0.956                   & 0.012                    & 0.721                   & 0.954                   & -0.003                   & 0.491                   & 0.972                  \\ \cline{2-12} 
\multicolumn{1}{c|}{}                                                                                   & \multirow{3}{*}{30} & $\eta_1^{(T)}$ & 0.044                    & 2.655                   & 0.948                   & -0.078                   & 1.691                   & 0.938                   & 0.055                    & 1.159                   & 0.940                  \\
\multicolumn{1}{c|}{}                                                                                   &                     & $\eta_2^{(T)}$ & -0.067                   & 1.610                   & 0.962                   & -0.002                   & 0.915                   & 0.964                   & -0.020                   & 0.651                   & 0.954                  \\
\multicolumn{1}{c|}{}                                                                                   &                     & $\eta_3^{(T)}$ & -0.067                   & 1.838                   & 0.958                   & -0.055                   & 1.122                   & 0.960                   & 0.051                    & 0.783                   & 0.946                  \\ \hline
\multicolumn{1}{c|}{\multirow{9}{*}{\begin{tabular}[c]{@{}c@{}}Independent\\ time points\end{tabular}}} & \multirow{3}{*}{10} & $\eta_1^{(T)}$ & -0.577                   & 0.373                   & 0.712                   & -0.558                   & 0.204                   & 0.370                   & -0.569                   & 0.131                   & 0.132                  \\
\multicolumn{1}{c|}{}                                                                                   &                     & $\eta_2^{(T)}$ & 0.199                    & 0.759                   & 0.962                   & 0.171                    & 0.447                   & 0.926                   & 0.175                    & 0.308                   & 0.888                  \\
\multicolumn{1}{c|}{}                                                                                   &                     & $\eta_3^{(T)}$ & -0.768                   & 1.472                   & 0.896                   & -0.762                   & 0.861                   & 0.866                   & -0.716                   & 0.656                   & 0.806                  \\ \cline{2-12} 
\multicolumn{1}{c|}{}                                                                                   & \multirow{3}{*}{20} & $\eta_1^{(T)}$ & 0.823                    & 1.301                   & 0.882                   & 0.844                    & 0.665                   & 0.724                   & 0.852                    & 0.441                   & 0.580                  \\
\multicolumn{1}{c|}{}                                                                                   &                     & $\eta_2^{(T)}$ & 6.158                    & 4.075                   & 0.636                   & 6.071                    & 2.479                   & 0.286                   & 6.075                    & 1.797                   & 0.066                  \\
\multicolumn{1}{c|}{}                                                                                   &                     & $\eta_3^{(T)}$ & 0.694                    & 3.095                   & 0.952                   & 0.665                    & 1.804                   & 0.936                   & 0.683                    & 1.315                   & 0.914                  \\ \cline{2-12} 
\multicolumn{1}{c|}{}                                                                                   & \multirow{3}{*}{30} & $\eta_1^{(T)}$ & 1.662                    & 1.491                   & 0.854                   & 1.626                    & 0.831                   & 0.598                   & 1.607                    & 0.530                   & 0.370                  \\
\multicolumn{1}{c|}{}                                                                                   &                     & $\eta_2^{(T)}$ & -10.111                  & 6.992                   & 0.700                   & -9.898                   & 4.801                   & 0.410                   & -9.989                   & 3.207                   & 0.126                  \\
\multicolumn{1}{c|}{}                                                                                   &                     & $\eta_3^{(T)}$ & -2.352                   & 1.880                   & 0.860                   & -2.446                   & 1.081                   & 0.374                   & -2.392                   & 0.715                   & 0.144                  \\ \hline
\multicolumn{1}{c|}{\multirow{9}{*}{\begin{tabular}[c]{@{}c@{}}Independent\\ mediators\end{tabular}}}   & \multirow{3}{*}{10} & $\eta_1^{(T)}$ & 0.000                    & 0.594                   & 0.972                   & 0.003                    & 0.351                   & 0.954                   & 0.008                    & 0.249                   & 0.964                  \\
\multicolumn{1}{c|}{}                                                                                   &                     & $\eta_2^{(T)}$ & -0.139                   & 0.640                   & 0.944                   & -0.175                   & 0.342                   & 0.944                   & -0.142                   & 0.263                   & 0.904                  \\
\multicolumn{1}{c|}{}                                                                                   &                     & $\eta_3^{(T)}$ & 0.164                    & 0.603                   & 0.942                   & 0.178                    & 0.368                   & 0.938                   & 0.172                    & 0.276                   & 0.892                  \\ \cline{2-12} 
\multicolumn{1}{c|}{}                                                                                   & \multirow{3}{*}{20} & $\eta_1^{(T)}$ & -0.045                   & 1.959                   & 0.948                   & 0.008                    & 1.215                   & 0.950                   & -0.023                   & 0.840                   & 0.952                  \\
\multicolumn{1}{c|}{}                                                                                   &                     & $\eta_2^{(T)}$ & -0.159                   & 1.370                   & 0.956                   & -0.135                   & 0.814                   & 0.950                   & -0.157                   & 0.576                   & 0.930                  \\
\multicolumn{1}{c|}{}                                                                                   &                     & $\eta_3^{(T)}$ & -0.764                   & 1.390                   & 0.926                   & -0.728                   & 0.830                   & 0.882                   & -0.779                   & 0.602                   & 0.772                  \\ \cline{2-12} 
\multicolumn{1}{c|}{}                                                                                   & \multirow{3}{*}{30} & $\eta_1^{(T)}$ & 0.044                    & 2.656                   & 0.946                   & -0.078                   & 1.691                   & 0.932                   & 0.055                    & 1.159                   & 0.948                  \\
\multicolumn{1}{c|}{}                                                                                   &                     & $\eta_2^{(T)}$ & -1.286                   & 2.087                   & 0.922                   & -1.154                   & 1.114                   & 0.872                   & -1.227                   & 0.864                   & 0.692                  \\
\multicolumn{1}{c|}{}                                                                                   &                     & $\eta_3^{(T)}$ & -0.741                   & 2.128                   & 0.934                   & -0.724                   & 1.279                   & 0.892                   & -0.632                   & 0.920                   & 0.884                  \\ \hline
\end{tabular}
\end{table}

For the finite-horizon setting, we consider the number of time points $T\in\{10,20,30\}$ and the sample size $n\in\{100,250,500\}$. Note that the true individual mediation effect cannot be derived analytically in our setting, and is computed numerically based on 10 million Monte Carlo samples. Table \ref{table:finite_sim} reports the results based on 500 replications. We observe that, when the number of subjects $n$ increases, the estimation bias and standard error both decrease for our proposed method, which agrees with our theory. {Moreover, our method achieves the desired coverage probability across all scenarios.} The same is not true for the two baseline methods. For instance, for the independent mediators method, the estimation bias of $\eta_2^{(T)}$ and $\eta_3^{(T)}$ does not decrease. This is because $M_{t1}$ is in the parent set of $M_{t2}$ and $M_{t3}$ for all $t=1,\dots,T$, and ignoring the effects along the paths $M_{t1}\to M_{t2}$ and $M_{t1}\to M_{t3}$ lead to a larger bias {and invalid coverage probability} in estimating $\eta_2^{(T)}$ and $\eta_3^{(T)}$. On the contrary, the estimation bias of $\eta_1^{(T)}$ is much smaller, since the parent set of $M_{t1}$ is empty given $(\bM_{t-1},R_t,A_t)$ in our simulation example.

\subsection{Infinite-horizon setting}
\label{ssec:sim_infinite}

\begin{table}[t!]
\centering
\caption{{Simulations for the infinite-horizon setting: the bias, the empirical standard error (SE) and the coverage probability (CP) between the estimated and true individual mediator effects with varying number of time points $T$ and number of subjects $n$.}}
\label{table:infinite_sim}
\begin{tabular}{ccc|rrr|rrr|rrr}
\hline
\multicolumn{3}{c|}{$n$}                                                                                                                             & \multicolumn{3}{c|}{20}                                                       & \multicolumn{3}{c|}{50}                                                       & \multicolumn{3}{c}{100}                                                      \\
Method                                                                                                  & $T$                  & Param               & \multicolumn{1}{c}{Bias} & \multicolumn{1}{c}{SE} & \multicolumn{1}{c|}{CP} & \multicolumn{1}{c}{Bias} & \multicolumn{1}{c}{SE} & \multicolumn{1}{c|}{CP} & \multicolumn{1}{c}{Bias} & \multicolumn{1}{c}{SE} & \multicolumn{1}{c}{CP} \\ \hline
\multicolumn{1}{c|}{\multirow{9}{*}{\begin{tabular}[c]{@{}c@{}}Proposed\\ method\end{tabular}}}         & \multirow{3}{*}{100} & $\eta_1^{(\infty)}$ & .002                    & .070                  & .918                     & .002                    & .046                  & .928                     & .001                    & .029                  & .934                    \\
\multicolumn{1}{c|}{}                                                                                   &                      & $\eta_2^{(\infty)}$ & .003                    & .045                  & .924                     & .001                    & .029                  & .932                     & .001                    & .020                  & .940                    \\
\multicolumn{1}{c|}{}                                                                                   &                      & $\eta_3^{(\infty)}$ & .003                    & .036                  & .924                     & .002                    & .021                  & .946                     & .000                    & .015                  & .946                    \\ \cline{2-12} 
\multicolumn{1}{c|}{}                                                                                   & \multirow{3}{*}{250} & $\eta_1^{(\infty)}$ & -.001                   & .042                  & .932                     & -.002                   & .026                  & .946                     & -.001                   & .019                  & .930                    \\
\multicolumn{1}{c|}{}                                                                                   &                      & $\eta_2^{(\infty)}$ & -.003                   & .029                  & .914                     & -.001                   & .019                  & .938                     & .001                    & .012                  & .958                    \\
\multicolumn{1}{c|}{}                                                                                   &                      & $\eta_3^{(\infty)}$ & .001                    & .020                  & .938                     & .000                    & .014                  & .938                     & -.001                   & .010                  & .950                    \\ \cline{2-12} 
\multicolumn{1}{c|}{}                                                                                   & \multirow{3}{*}{500} & $\eta_1^{(\infty)}$ & .000                    & .029                  & .944                     & .000                    & .018                  & .956                     & -.001                   & .014                  & .932                    \\
\multicolumn{1}{c|}{}                                                                                   &                      & $\eta_2^{(\infty)}$ & .000                    & .020                  & .928                     & -.001                   & .012                  & .954                     & .000                    & .009                  & .942                    \\
\multicolumn{1}{c|}{}                                                                                   &                      & $\eta_3^{(\infty)}$ & .000                    & .016                  & .924                     & -.001                   & .010                  & .932                     & .000                    & .007                  & .928                    \\ \hline
\multicolumn{1}{c|}{\multirow{9}{*}{\begin{tabular}[c]{@{}c@{}}Independent\\ time points\end{tabular}}} & \multirow{3}{*}{100} & $\eta_1^{(\infty)}$ & .371                    & .047                  & .000                     & .370                    & .030                  & .000                     & .371                    & .021                  & .000                    \\
\multicolumn{1}{c|}{}                                                                                   &                      & $\eta_2^{(\infty)}$ & -.065                   & .076                  & .818                     & -.070                   & .049                  & .658                     & -.071                   & .034                  & .404                    \\
\multicolumn{1}{c|}{}                                                                                   &                      & $\eta_3^{(\infty)}$ & .132                    & .086                  & .628                     & .133                    & .060                  & .314                     & .132                    & .040                  & .076                    \\ \cline{2-12} 
\multicolumn{1}{c|}{}                                                                                   & \multirow{3}{*}{250} & $\eta_1^{(\infty)}$ & .372                    & .027                  & .000                     & .371                    & .018                  & .000                     & .371                    & .013                  & .000                    \\
\multicolumn{1}{c|}{}                                                                                   &                      & $\eta_2^{(\infty)}$ & -.075                   & .048                  & .616                     & -.073                   & .030                  & .300                     & -.073                   & .020                  & .052                    \\
\multicolumn{1}{c|}{}                                                                                   &                      & $\eta_3^{(\infty)}$ & .134                    & .055                  & .286                     & .135                    & .034                  & .028                     & .133                    & .024                  & .000                    \\ \cline{2-12} 
\multicolumn{1}{c|}{}                                                                                   & \multirow{3}{*}{500} & $\eta_1^{(\infty)}$ & .370                    & .021                  & .000                     & .371                    & .013                  & .000                     & .371                    & .010                  & .000                    \\
\multicolumn{1}{c|}{}                                                                                   &                      & $\eta_2^{(\infty)}$ & -.073                   & .035                  & .368                     & -.074                   & .021                  & .058                     & -.073                   & .015                  & .002                    \\
\multicolumn{1}{c|}{}                                                                                   &                      & $\eta_3^{(\infty)}$ & .131                    & .039                  & .074                     & .131                    & .024                  & .000                     & .134                    & .018                  & .000                    \\ \hline
\multicolumn{1}{c|}{\multirow{9}{*}{\begin{tabular}[c]{@{}c@{}}Independent\\ mediators\end{tabular}}}   & \multirow{3}{*}{100} & $\eta_1^{(\infty)}$ & .002                    & .070                  & .918                     & .002                    & .046                  & .934                     & .001                    & .029                  & .936                    \\
\multicolumn{1}{c|}{}                                                                                   &                      & $\eta_2^{(\infty)}$ & .322                    & .031                  & .000                     & .324                    & .019                  & .000                     & .325                    & .015                  & .000                    \\
\multicolumn{1}{c|}{}                                                                                   &                      & $\eta_3^{(\infty)}$ & -.264                   & .038                  & .000                     & -.266                   & .024                  & .000                     & -.265                   & .016                  & .000                    \\ \cline{2-12} 
\multicolumn{1}{c|}{}                                                                                   & \multirow{3}{*}{250} & $\eta_1^{(\infty)}$ & -.001                   & .042                  & .934                     & -.002                   & .026                  & .934                     & -.001                   & .019                  & .942                    \\
\multicolumn{1}{c|}{}                                                                                   &                      & $\eta_2^{(\infty)}$ & .324                    & .019                  & .000                     & .324                    & .012                  & .000                     & .325                    & .009                  & .000                    \\
\multicolumn{1}{c|}{}                                                                                   &                      & $\eta_3^{(\infty)}$ & -.265                   & .024                  & .000                     & -.265                   & .014                  & .000                     & -.265                   & .010                  & .000                    \\ \cline{2-12} 
\multicolumn{1}{c|}{}                                                                                   & \multirow{3}{*}{500} & $\eta_1^{(\infty)}$ & .000                    & .029                  & .936                     & .000                    & .018                  & .952                     & -.001                   & .014                  & .936                    \\
\multicolumn{1}{c|}{}                                                                                   &                      & $\eta_2^{(\infty)}$ & .324                    & .014                  & .000                     & .325                    & .009                  & .000                     & .324                    & .006                  & .000                    \\
\multicolumn{1}{c|}{}                                                                                   &                      & $\eta_3^{(\infty)}$ & -.265                   & .017                  & .000                     & -.265                   & .011                  & .000                     & -.265                   & .007                  & .000                    \\ \hline
\end{tabular}
\end{table}

For the infinite-horizon setting, we consider the number of time points $T\in\{100,250,500\}$ and the sample size $n\in\{20,50,100\}$. Again, the true individual mediation effect cannot be derived analytically in our setting, and is computed numerically based on 10 million Monte Carlo samples with $3,000$ time points. We also drop the first five steps as a warm up to mitigate the potential influence of the initial conditions. Table \ref{table:infinite_sim} reports the results based on 500 data replications. We again observe that the estimation bias and standard error both decrease for our method as $n$ increases, {and the desired coverage probability is achieved, across all scenarios.} On the other hand, the bias increases sharply for the independent time points method, because ignoring the carryover effects becomes exaggerated with a large $T$. Moreover, the estimation bias of the independent mediators method is much larger than the proposed method for $\eta_2^{(\infty)}$ and $\eta_3^{(\infty)}$. {Both baseline methods fail to achieve the desired coverage probability in most cases, except for the independent mediators method with $\eta_1^{(\infty)}$. This is due to the specific simulation setting with an empty parent set given the history.} In general, these results demonstrate the importance of taking into account both the dependence among multiple time points and among multiple mediators.

\section{Data Application}
\label{sec:real}

In this section, we revisit the motivating Intern Health Study \citep{necamp2020assessing}. The IHS is a 26-week sequentially randomized trial with the objective of understanding the biological mechanisms of depression, and the ultimate goal of improving the mental health outcomes of medical interns in the United States. The study developed and deployed a mobile app to deliver prompt notifications, such as reminders to have a break, take a walk, or prioritize sleep, which aims to improve the well-being of interns who often work under stressful environments. In each week, each intern was randomized into receiving the notifications or not. Meanwhile, wearable devices (Fitbit) recorded daily measurements of step count (Step), sleep duration (Sleep, minutes), resting heart rate (RHR, beats per minute), and heart rate variability (HRV, milliseconds). Interns also self-reported a daily mood score in the app \citep{shaffer2017overview}. We formulate the problem in the framework of multivariate dynamic mediation analysis, with the binary status of receiving the notification or not as the treatment, the transformed measurements, i.e., the cubic-root of step count, the square-root of sleep duration, RHR and HRV as $d=4$ potential mediators, and the mood score as the outcome. We average all the measurements within each week, resulting in $T=26$ weeks of data, for $n=1196$ interns undergoing the sequential randomization.

\begin{figure}[t!]
\centering
\includegraphics[width=0.95\linewidth]{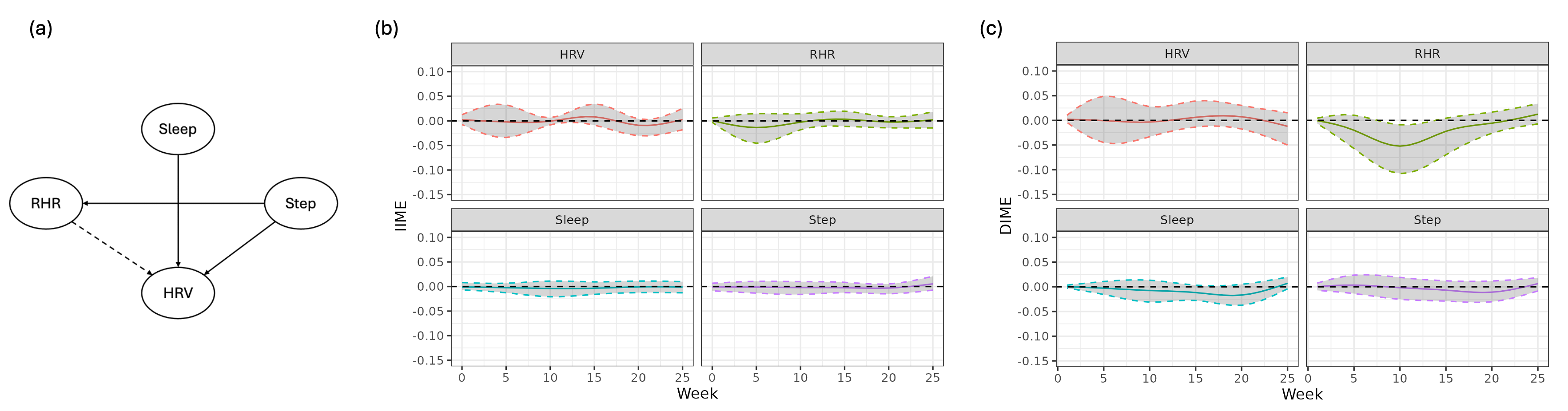} 
\caption{{Analysis of IHS mobile health data: (a) the estimated DAG among the four mediators, dashed/solid arrows represent negative/positive effects; (b) the estimated immediate individual mediation effects over time; (c) the estimated delayed individual mediation effects over time. The shaded area indicates the $95\%$ confidence interval.}}
\label{fig:ihs2018}
\end{figure}

\begin{figure}[t!]
\centering
\includegraphics[width=0.6\linewidth,height=1.5in]{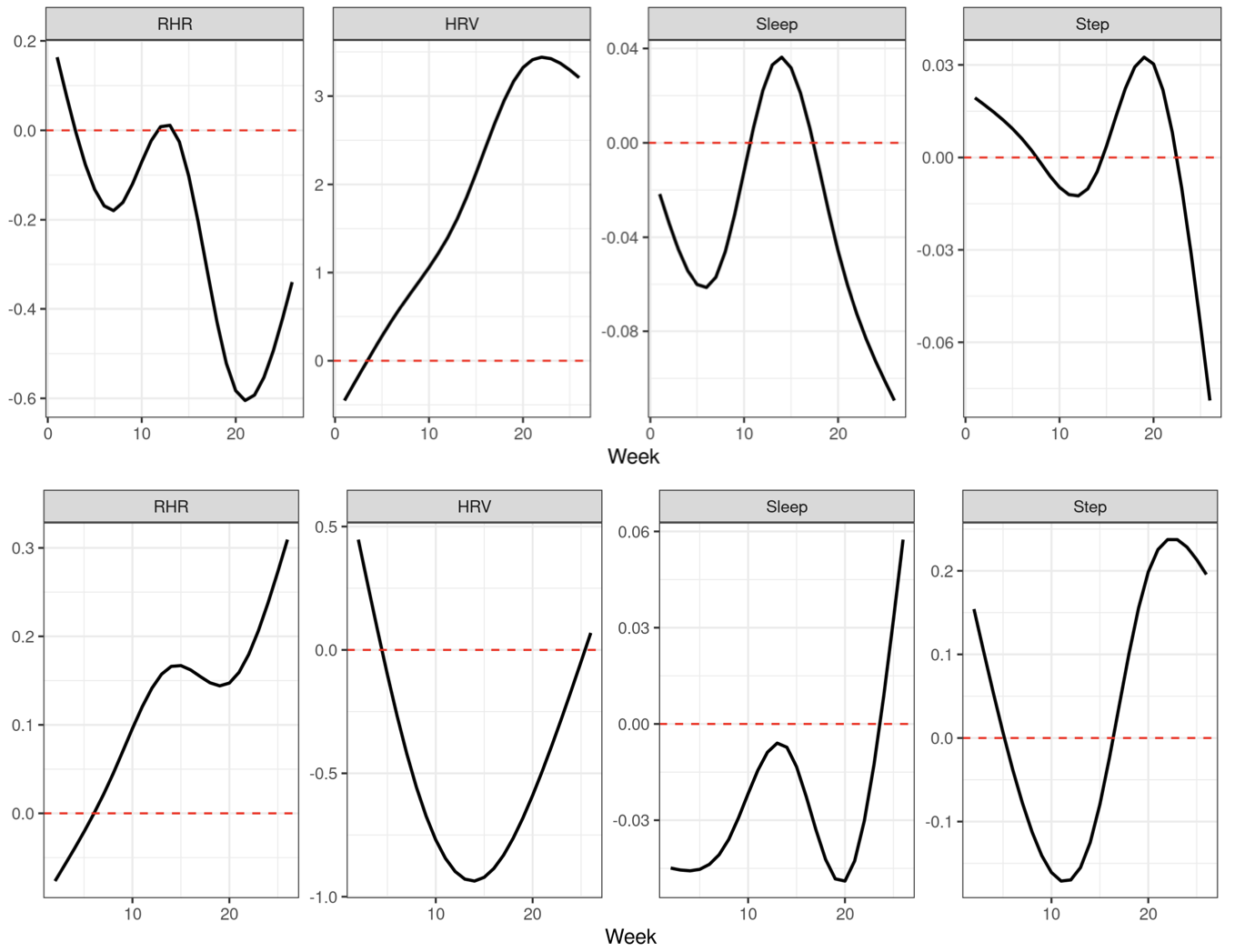} 
\caption{Analysis of IHS mobile health data: the estimated effects from the treatment to the four individual mediators, decomposed as the immediate (first row) and delayed (second row) effects, respectively.}
\label{fig:ihs2018-A2M}
\end{figure}

{We apply our method to this data adopting the finite-horizon setting. This is partly because we have a relatively limited number of time points $T=26$ in this study, and partly because the data is likely to be non-stationary over time \citep{necamp2020assessing,li2022testing,wang2023robust}. Our analysis, as shown in Figure \ref{fig:ihs2018-A2M}, also confirms that the individual mediation effect varies over time.} 
Figure \ref{fig:ihs2018}(a) shows the estimated DAG structure among the four mediators, while Figures \ref{fig:ihs2018}(b) and (c) show the estimated individual mediation effects over time, along with the associated $95\%$ confidence intervals, that are smoothed with a natural cubic spline and further decomposed as the immediate and delayed effects. 

We make a number of observations. First of all, we see from Figure \ref{fig:ihs2018}(a) that, the four mediators have complex relationships between each other, indicating the importance of accounting for the dependence structure among the mediators. Second, we see from Figure \ref{fig:ihs2018}(b) and (c) that, the magnitude of the delayed effects is generally greater than that of the immediate effects, indicating the importance of accounting for the carryover effects in our understanding of the effects of push notification to intern's mood. {Third, most individual mediation effects are insignificant. This could potentially be attributed to the relatively weak treatment effect, as also noted by \cite{necamp2020assessing}.} Fourth, there is a significantly negative carryover effect from the push notification to mood score mediated by the RHR around week $10$. Meanwhile, this carryover effect is predominantly negative across most weeks. To illustrate this effect, Figure \ref{fig:ihs2018-A2M} further plots the smoothed estimated effects from the treatment to the four individual mediators, decomposed as the immediate and delayed effects, respectively. We see that the push notification leads to an increased RHR after week five, suggesting that the push notification can lead to a lower mood score by increasing the RHR. This agrees with our prior knowledge that an increased RHR usually indicates a more tired and stressful state, thus a worse mood. {Finally, the carryover effect from the push notification to sleep duration is mostly negative, and is nearly significant between weeks 15 and 20. As seen in Figure \ref{fig:ihs2018-A2M}, the push notification generally results in reduced sleep hours, aligning with our knowledge that fewer sleep hours lead to poorer mood too.}

{\section{Conclusion and Discussion}
\label{sec:discuss}
This paper proposes a reinforcement learning framework for multivariate dynamic mediation analysis where there are multivariate and conditionally dependent mediators observed over multiple time points. While most existing studies focus on individual mediation effects among multivariate mediators or the effects of a single set of mediators over time, they typically do not consider both. A recent study by \citet{wei2024time} explores individual mediation effects across multiple time points. However, it does not account for delayed mediation effects, nor does it analyze the indirect effects of each mediator from its upstream mediators. 

Below, we discuss the relation between our individual mediation effect and the Granger causality, numerous other mediation effects that may be of potential interest, and some additional extensions.

\subsection{Connection to Granger causality}
\label{ssec:granger}

The Granger causality is a concept that leverages the temporal ordering inherent to time series so to draw causal statements restricted to the ``past'' causing the ``future'' \citep{granger1969investigating}. A variable is said to Granger-cause the other, if including this variable's past values provides valuable information for forecasting the other variable's future behavior. The proposed models in \eqref{eq:LSEM} and \eqref{eq:model_setup} are mathematically similar to those under the Granger causality framework, since $R_t$ may be predicted by its past value $R_{t-1}$, $\bm{M}_t$ and $\bm{M}_{t-1}$, and $\bm{M}_t$ also depends on its past value $\bm{M}_{t-1}$ and $R_{t-1}$.

\begin{figure}[t]
\centering
\includegraphics[width=0.6\linewidth]{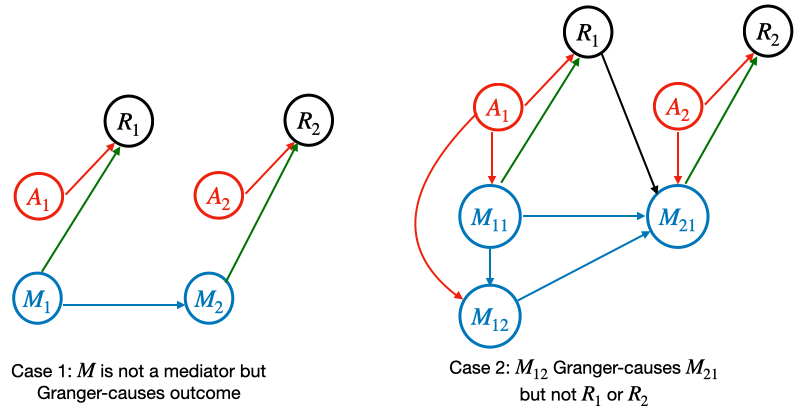}
\caption{An illustration of the difference between the individual mediation effect and the Granger causality.}
\label{fig:Granger}
\end{figure}
 
However, our notion of individual mediation effect differs from the concept of Granger causality, in that a mediator's individual mediation effect and its ability to Granger-cause the outcome is \emph{not} equivalent. On the one hand, a mediator can have a zero individual mediation effect, but still Granger-causes the outcome, if it is not influenced by the treatment. Figure~\ref{fig:Granger}, Case 1, gives an illustrative example. In this case, we have a single mediator that impacts the outcome at each time point, thereby Granger-causing the outcome. However, since the mediator is unaffected by the treatment, there is no direct causal pathway from the treatment through the mediator to the outcome, resulting in a zero individual mediation effect.  On the other hand, a mediator with a nonzero individual mediation effect might not necessarily Granger-cause the outcome, if it does not improve the forecast accuracy of the outcome beyond other mediators. Figure~\ref{fig:Granger}, Case 2, gives another example. In this case, the first set of mediators $M_{11}$ and $M_{21}$ directly influence the outcome, thus Granger-causing the outcome. The second mediator $M_{12}$, however, affects the outcome indirectly through its influence on the first mediator $M_{21}$, resulting in a nonzero individual mediation effect. Despite this, $M_{12}$ does not Granger-cause the outcome $R_1$ or $R_2$, as its influence is mediated indirectly.

\subsection{Sequential mediation effect}
\label{ssec:sequential}

As discussed in Section \ref{ssec:definition}, the individual mediation effect $\eta_j^{(T)}$ measures the cumulative effect mediated through the $j$th mediator $M_j$ across all the upstream mediators and all $T$ time points. Meanwhile, it may be of separate interest to analyze the portion of the individual mediation effect of $M_j$ that is attributed to its upstream mediator $M_k$. We denote this effect as $\eta_{k\to j}$, and call it the \emph{sequential} mediation effect, as it quantifies the effect that is sequentially transmitted from the treatment, through the $k$th mediator, to the $j$th mediator, then to the outcome. 

We next formally define the sequential mediation effect. Toward that end, we require the DAG structure to remain stationary over time. That is, if the $k$th mediator is an ancestor of the $j$th mediator at any time $t$, i.e., there exists a direct path from $M_{tk}$ to $M_{tj}$, then the same relation holds at all time points. For the finite-horizon setting, we define the sequential mediation effect over $T$ time points as
\begin{eqnarray*}\label{eqn:sequentialeffect}
 \begin{split}
&\eta_{k\to j}^{(T)} =\sum_{t=1}^T\bigg\{\frac{\partial}{\partial a}\mathbb{E}[R_t\mid do(A_s=a, \forall s\leq t)] - \frac{\partial}{\partial a}\mathbb{E}[R_t\mid do(A_s=a, M_{sj}=m, \forall s\leq t)] \\
-& \frac{\partial}{\partial a}\mathbb{E}[R_t\mid do(A_s=a, M_{sk}=m, \forall s\leq t)] + \frac{\partial}{\partial a}\mathbb{E}[R_t\mid do(A_s=a, M_{sj}=M_{kj}=m, \forall s\leq t)]\bigg\},
\end{split}
\end{eqnarray*}
if the $k$th mediator is an ancestor of the $j$th mediator, and set $\eta_{k\to j}^{(T)} = 0$ otherwise. In this definition, the first term corresponds to the total treatment effect, the second and third terms measure the effects that do not go through the $j$th or the $k$th mediator, respectively, and the last term quantifies the effect that does not simultaneously pass both the $j$th and $k$th mediators. Then, by the principle of inclusion-exclusion, $\eta_{k\to j}^{(T)}$ measures the desired sequential mediation effect. For the infinite-horizon setting, we define $\eta_{k\to j}^{(\infty)} = \lim_{T\to \infty}\eta_{k\to j}^{(T)}/T$, provided the limit exists.

To further illustrate this definition, we revisit the examples in Figure~\ref{fig:two-stage}. For the single-stage example in Figure~\ref{fig:two-stage}, left panel, the mediator $M_{12}$ is a child of $M_{11}$ and is affected by $M_{11}$. Its individual mediation effect that passes through $M_{11}$ is 
\begin{eqnarray*}
  \eta_{1 \to 2}^{(1)} & = & \frac{\partial}{\partial a}\mathbb{E}[R_1\mid do(A_1=a)] - \frac{\partial}{\partial a}\mathbb{E}[R_1\mid do(A_1=a, M_{12}=m)] \\
  & & - \
  \frac{\partial}{\partial a}\mathbb{E}[R_1\mid do(A_1=a, M_{11}=m)] + \frac{\partial}{\partial a}\mathbb{E}[R_1\mid do(A_1=a, M_{11}=M_{12}=m)] \\
  & = & \theta_{A_1\to R_1} - \theta_{A_1\to R_1}^{M_{12}} - \theta_{A_1\to R_1}^{M_{11}} + \theta_{A_1\to R_1}^{M_{11},M_{12}} \\
  & = & \theta_{A_1\to M_{11}}\theta_{M_{11}\to M_{12}}\theta_{M_{12}\to R_1}.
\end{eqnarray*}
It is clear to see that the term $\theta_{A_1\to M_{11}}\theta_{M_{11}\to M_{12}}\theta_{M_{12}\to R_1}$ corresponds to the effect along the path $A_1\to M_{11}\to M_{12}\to R_1$, which coincides with the results from the path analysis. Similarly, for the two-stage example in Figure~\ref{fig:two-stage}, right panel, we have that, 
\begin{eqnarray*}
\begin{split}
\eta_{1 \to 2}^{(2)}&=\eta_{1 \to 2}^{(1)} + \frac{\partial}{\partial a}\mathbb{E}[R_2\mid do(A_1=A_2=a)]-\frac{\partial}{\partial a}\mathbb{E}[R_2\mid do(A_1=A_2=a,M_{11}=M_{21}=m)] \\
  &\hspace{2em}- \frac{\partial}{\partial a}\mathbb{E}[R_2\mid do(A_1=A_2=a,M_{12}=M_{22}=m)] \\
  &\hspace{2em}+ \frac{\partial}{\partial a}\mathbb{E}[R_2\mid do(A_1=A_2=a,M_{11}=M_{21}=M_{12}=M_{22}=m)] \\
  &=\eta_{1 \to 2}^{(1)} + \theta_{A_1,A_2\to R_2}-\theta_{A_1,A_2\to R_2}^{M_{11},M_{21}}
  - \theta_{A_1,A_2\to R_2}^{M_{12},M_{22}} + \theta_{A_1,A_2\to R_2}^{M_{11},M_{12},M_{21},M_{22}} \\
  &=\eta_{1 \to 2}^{(1)} + \theta_{A_1\to R_2} -\theta_{A_1\to R_2}^{M_{11},M_{21}}  - \theta_{A_1\to R_2}^{M_{12},M_{22}} +\theta_{A_1\to R_2}^{M_{11},M_{21},M_{12},M_{22}}\\ &\hspace{2em}+\theta_{A_2\to R_2}-\theta_{A_2\to R_2}^{M_{21}} - \theta_{A_2\to R_2}^{M_{22}} +\theta_{A_2\to R_2}^{M_{21},M_{22}},
\end{split}
\end{eqnarray*}
where the terms in the last two lines of the above equation can be calculated recursively with the formulas in equations~\eqref{eqn:recursive-A} and~\eqref{eqn:recursive-M}, allowing for the estimation and inference of the sequential mediation effect.

Finally, we extend beyond the above illustrative examples to outline the estimation and inference procedure for the sequential mediation effect under a general setting involving $d$ mediators across $T$ time points. As for estimation, we begin by applying an existing structural learning algorithm to the observed data to learn the DAG structure. From the estimated DAG, if the $k$th mediator is not an ancestor of the $j$th mediator, we set the estimated sequential effect to zero. Otherwise, we obtain the estimated $\widehat{\eta}_{k\to j}$ through the quantities $\theta_{A_1,\ldots,A_t\to R_t}$, $\theta_{A_1,\ldots,A_t\to R_t}^{M_{1j},\ldots,M_{tj}}$, and $\theta_{A_1,\cdots,A_t\to R_t}^{M_{1k},M_{1j},\ldots,M_{kj},M_{tj}}$, which in turn can be recursively updated following the method developed in Section \ref{sec:method}. As for inference, we again apply the bootstrap approach in Section \ref{sec:method} to quantify the uncertainty of this plug-in estimator $\widehat{\eta}_{k\to j}$.

\subsection{Direct, indirect and conditional effects}
\label{ssec:moreeffects}

We discuss the direct effect DE and indirect effect IE in Section \ref{ssec:totaleffect}. We now briefly discuss their estimation and inference, which may be of separate interest. Recall DE in \eqref{eqn:controlledDE} under linear SEMs. It satisfies a recursive equation,
\begin{equation*}
\text{DE}^{(t)} = \text{DE}^{(t-1)} \ \zeta_{2t} + \delta_{2t},
\end{equation*}
for $t=2,\ldots,T$, with $\text{DE}^{(1)} = \delta_{21}$. Here we add the superscript $(t)$ to DE to indicate the direct effect across $t$ time points. This recursive formulation allows us to employ a similar plug-in method as in Section \ref{sec:method} to estimate DE, as well as a similar bootstrap approach for inference. To estimate IE, we first compute the total effect as a byproduct of the proposed procedure in Section \ref{sec:method}. We then estimate IE as the difference between the estimated total effect and DE. Again, we can employ the bootstrap approach to infer IE. 

Our definition of the individual mediation effect is concerned with the \emph{marginal} effect, 
\begin{equation*}
\frac{\partial}{\partial a}\mathbb{E}[R_t\mid do(A_t=a)] - \frac{\partial}{\partial a}\mathbb{E}[R_t\mid do(A_t=a, M_{tj}=m)], 
\end{equation*}
which marginalizes over the history $\{A_s, \bM_s, R_s \}$ for all $s<t$. This marginalization is appropriate because our interventions on the treatment and mediator at time $t$ do not causally influence those historical variables. Henceforth, the distribution of those historical variables remains unchanged before and after the intervention. Meanwhile, it may be of separate interest to study the \emph{conditional} effect, 
\begin{align*}
\frac{\partial}{\partial a}\mathbb{E}[R_t\mid do(A_t=a), & \{A_s, \bM_s, R_s \},\forall s<t] \\ 
& - \frac{\partial}{\partial a}\mathbb{E}[R_t\mid do(A_t=a, M_{tj}=m),\{A_s, \bM_s, R_s \},\forall s<t],
\end{align*}
given the history. Nevertheless, we note that, under the linear structural model, the conditional effect is equivalent to the marginal effect that we target in this article. We next sketch a few lines to show the equivalence. Specifically, the first term is $\frac{\partial}{\partial a}\mathbb{E}[R_t\mid do(A_t=a)]=\theta_{A_t\to R_t}=\beta_{A_t, R_t|\{A_s, \bM_s, R_s \},\forall s<t}$ in our current definition. It is the partial coefficient of $A_t$ by linearly regressing $R_t$ on $A_t$ given the historical variables that could affect the distribution of $A_t$. Under the linear model structures \eqref{eq:LSEM} and \eqref{eq:model_setup}, the effects of $A_t$ and the historical variables on $R_t$ are additive. Consequently, the resulting conditional effect $\frac{\partial}{\partial a}\mathbb{E}[R_t\mid do(A_t=a),\{A_s, \bM_s, R_s \},\forall s<t] = \beta_{A_t,R_t\mid \{A_s, \bM_s, R_s \},\forall s<t}$ remains the same as the marginal effect, regardless of the values of the historical variables. The second term is $\frac{\partial}{\partial a}\mathbb{E}[R_t\mid do(A_t=a, M_{tj}=m)]=\theta_{A_t\to R_t}^{M_{tj}} = \theta_{A_t\to R_t} - \theta_{A_t\to M_{tj}}\theta_{M_{tj}\to R_t}$. Again, the value $\theta_{A_t\to R_t}$ remains the same whether or not conditioning on the historical variables. Similarly, under \eqref{eq:LSEM} and \eqref{eq:model_setup}, the effects of $M_{tj}$ and the historical variables on $R_t$ are additive. Consequently, the value of the product $\theta_{A_t\to M_{tj}}\theta_{M_{tj}\to R_t}$ remains the same, regardless of whether the history is included in the conditioning set or not.

\subsection{Extension to varying $T$ and time lags}
\label{ssec:varyingT}

Finally, we remark that our method can be adapted to accommodate the setting when the number of time points $T$ varies among subjects, and when the time lags between two time points differ. 

For the finite-horizon setting, $T$ can be different for different subjects, because the model parameters and intermediate quantities at any given time $t$ are estimated by pooling the data across subjects who have observations available at that time point. Besides, Theorem \ref{theory:finite} remains valid when $T$ varies across subjects, provided that the sample size at each time point is sufficiently large. Meanwhile, for the time lag between two time points, we require it to follow the same distribution across all subjects, but it does not need to follow the same distribution across all time points, since we allow the data generating process to be non-stationary over time.

For the infinite-horizon setting, $T$ again can be different, because the data are pooled across both subjects and time points. Similarly, Theorem \ref{theory:infinite} remains valid when $T$ varies across subjects, provided that the total sample size is sufficiently large. Meanwhile, the time lag between two time points needs to follow the same distribution, due to the requirement for stationarity for the infinite-horizon setting. However, this requirement does not extend to between subjects, and the distributions can vary among different subjects. In that case, we can use each subject's own data to estimate their subject-specific mediation effects.
}

\begin{acks}[Acknowledgments]
The authors are grateful for the contributions of the researchers, administrators, and participants involved in the Intern Health Study (\url{https://clinicaltrials.gov/study/NCT03972293}). The authors also thank the editor, associate editor and reviewers for their comments, which have led to substantial improvements of the manuscript. 
\end{acks}
\begin{funding}
	Luo's research was partly supported by NIH grant R21AG083364. Shi's research was supported in part by the EPSRC grant EP/W014971/1. Wu's work was partly supported by NIH grants R01 MH101459 and R01 NR013658. Li's research was partially supported by NSF grant CIF-2102227, NIH grants R01AG062542 and R01AG080043.
%
\end{funding}

\begin{supplement}
The supplement provides the proofs of the main theorems in the paper.
\end{supplement}

\bibliography{ref-mediation}
\bibliographystyle{imsart-nameyear}

\appendix

\subsection{Proof of Theorem \ref{thm:eta_finite}}
\label{ssec:proof_finite}

By Definition~\ref{def:finite}, the individual mediation effect of the $j$th mediator, $j=1,\ldots,d$, is of the form, 
\begin{eqnarray}  \label{eq:eta_finite_stage}
\eta_j^{(t)} & = & \sum_{s=1}^t\frac{\partial}{\partial a}\mathbb{E}[R_s\mid do(A_i=a,\forall i \leq s)] - \sum_{s=1}^t\frac{\partial}{\partial a}\mathbb{E}[R_s\mid do(A_i=a, M_{ij}=m,\forall i\leq s)] \nonumber \\
& = & (t-1) \eta_j^{(t-1)} + \left\{\frac{\partial}{\partial a}\mathbb{E}[R_t\mid do(A_i=a,\forall i \leq t)] \right. \nonumber \\
& & \hspace{1.5in} \left. - \frac{\partial}{\partial a}\mathbb{E}[R_t\mid do(A_i=a,M_{ij}=m,\forall i \leq t)] \right\} \nonumber \\
& = & (t-1)\eta_j^{(t-1)} + \left(\theta_{A_1,\ldots,A_t\to R_t} - \theta_{A_1,\ldots,A_t\to R_t}^{M_{1j},\ldots,M_{tj}} \right).
\end{eqnarray}
The key is to express \eqref{eq:eta_finite_stage} as a function of the intermediate quantities in Section~\ref{ssec:intermediate} of the paper. 

First, because all the treatments $(A_1,\ldots,A_t)$ are randomly assigned, and thus are independent of each other and all other covariates, we have
\begin{eqnarray*}
\theta_{A_1,\ldots, A_t\to R_t} & = & \theta_{A_1\to R_t} +\theta_{A_2\to R_t} + \ldots + \theta_{A_t\to R_t}, \\
\theta_{A_1,\ldots,A_t\to R_t}^{M_{1j},\ldots,M_{tj}} & = & \theta_{A_1\to R_t}^{M_{1j},\ldots,M_{tj}} + \theta_{A_2\to R_t}^{M_{2j},\ldots,M_{tj}} + \ldots + \theta_{A_t\to R_t}^{M_{tj}}.
\end{eqnarray*}

Next, we expand the terms in the summation of the second equation. Note that, when $s=t$, $\theta_{A_t\to R_t}^{M_{tj}}=\theta_{A_t\to R_t} - \theta_{A_t\to M_{tj}}\theta_{M_{tj}\to R_t}$. For any $s=1,\ldots,t-1$, the effect of $A_s$ to $R_t$ in a joint intervention on $(M_{sj}, M_{(s+1)j}\ldots,M_{tj})$ is
\begin{equation}\label{eq:finite_single_term}
\begin{split}
\theta_{A_s\to R_t}^{M_{sj},\ldots,M_{tj}} &=
\theta_{A_s\to R_t}^{M_{(s+1)j},\ldots,M_{tj}}-\theta_{A_s\to M_{sj}}\theta_{M_{sj}\to R_t}^{M_{(s+1)j},\ldots, M_{tj}}\\
&=\theta_{A_s\to R_t}^{M_{(s+2)j},\ldots,M_{tj}}-\theta_{A_s\to M_{(s+1)j}}\theta_{M_{(s+1)j}\to R_t}^{M_{(s+1)j},\ldots,M_{tj}}-\theta_{A_s\to M_{sj}}\theta_{M_{sj}\to R_t}^{M_{sj},\ldots, M_{tj}} \\
& = \ldots \\
&=\theta_{A_s\to R_t}  - \sum_{i=s}^{t}\theta_{A_s\to M_{ij}}\theta_{M_{ij}\to R_t}^{M_{ij},\ldots,M_{tj}}.
\end{split}
\end{equation}

Next, we derive the recursive formula for computing $\theta_{M_{sj}\to R_t}^{M_{sj},\ldots,M_{tj}}$, with $s=1,\ldots,t-1$. The idea is similar to \eqref{eq:finite_single_term}, i.e., 
\begin{equation}\label{eq:finite_single_term_M}
\begin{split}
	\theta_{M_{sj}\to R_t}^{M_{sj},\ldots,M_{tj}} & = \theta_{M_{sj}\to R_t}^{M_{(s+2)j},\ldots, M_{tj}} - \theta_{M_{sj}\to M_{(s+1)j}}\theta_{M_{(s+1)j}\to R_t}^{M_{(s+1)j},\ldots,M_{tj}} \\
	& = \theta_{M_{sj}\to R_t}^{M_{(s+3)j},\ldots,M_{tj}} - \theta_{M_{sj}\to M_{(s+2)j}} \theta_{M_{(s+2)j}\to R_t}^{M_{(s+2)j},\ldots, M_{tj}} - \theta_{M_{sj}\to M_{(s+1)j}}\theta_{M_{(s+1)j}\to R_t}^{M_{(s+1)j},\ldots,M_{tj}} \\
	&=\ldots \\
	& = \theta_{M_{sj}\to R_t} - \sum_{i=s+1}^{t} \theta_{M_{sj}\to M_{ij}}\theta_{M_{ij}\to R_t}^{M_{ij},\ldots,M_{tj}}.
\end{split}
\end{equation}
Given $\theta_{M_{ij}\to R_t}^{M_{ij},\ldots,M_{tj}}$ for $i=s+1,\ldots,t$, the above recursion allows us to evaluate $\theta_{M_{sj}\to R_t}^{M_{sj},\ldots,M_{tj}}$.

Given the expanded forms in \eqref{eq:finite_single_term} and~\eqref{eq:finite_single_term_M}, we compute the total effects from $(A_1,\ldots,A_t)$ on $R_t$ in a joint intervention on $(M_{1j},\ldots,M_{tj})$ as,
\begin{eqnarray*}
\theta_{A_1,\ldots,A_t\to R_t}^{M_{1j},\ldots,M_{tj}} 
& = & \theta_{A_1\to R_t}^{M_{1j},\ldots,M_{tj}} + \theta_{A_2\to R_t}^{M_{2j},\ldots,M_{tj}} + \ldots + \theta_{A_t\to R_t}^{M_{tj}} \\
& = & \left(\theta_{A_1\to R_t}- \sum_{i=1}^{t}\theta_{A_1\to M_{ij}}\theta_{M_{ij}\to R_t}^{M_{ij},\ldots,M_{tj}} \right) + \left(\theta_{A_2\to R_t}-\sum_{i=2}^{t}\theta_{A_2\to M_{ij}}\theta_{M_{ij}\to R_t}^{M_{ij},\ldots,M_{tj}}  \right)\\
& & \hspace{2em} +\ldots+\left(\theta_{A_{t}\to R_t}-\theta_{A_t\to M_{tj}}\theta_{M_{ij}\to R_t}^{M_{ij},\ldots,M_{tj}}  \right) \\
& = & \sum_{s=1}^t \theta_{A_s\to R_t} -\sum_{s=1}^t\sum_{i=s}^t \theta_{A_s\to M_{ij}} \theta_{M_{ij}\to R_t}^{M_{ij},\ldots,M_{tj}} ,
\end{eqnarray*}
where $\theta_{M_{ij}\to R_t}^{M_{ij},\ldots,M_{tj}}$ can be calculated via \eqref{eq:finite_single_term_M}.

Finally, we plug in the above expansion to \eqref{eq:eta_finite_stage}, and obtain that, 
\begin{equation*}
\eta_j^{(t)} =  \eta_j^{(t-1)} +  \sum_{s=1}^t\sum_{i=s}^t \theta_{A_s\to M_{ij}} \theta_{M_{ij}\to R_t}^{M_{ij},\ldots,M_{tj}} , \ t=1,\ldots,T.
\end{equation*}
This completes the proof of Theorem \ref{thm:eta_finite}.
\eop

\subsection{Proof of Theorem \ref{thm:eta_infinite}}
\label{ssec:proof_infinite}

First, by the method of induction, we have, 
\begin{eqnarray*}
\eta_j^{(1)} & = & \theta_{M_{1j}\to R_1}\theta_{A_1\to M_{1j}}, \\
\eta_j^{(2)} & = &  \theta_{M_{1j}\to R_1} (2\theta_{A_1\to M_{1j}} + \theta_{A_1\to M_{2j}}) + \theta_{M_{1j}\to R_2}^{M_{1j},M_{2j}} \theta_{A_1\to M_{1j}}, \\
\eta_j^{(3)} & = & \theta_{M_{1j}\to R_1} (3 \theta_{A_1\to M_{1j}}+ 2 \theta_{A_1\to M_{2j}} + \theta_{A_1\to M_{1j}}) + \theta_{M_{1j}\to R_3}^{M_{1j},M_{2j},M_{3j}}  \theta_{A_1\to M_{1j}}  \\
& & \quad\quad\quad  + \; \theta_{M_{1j}\to R_2}^{M_{1j},M_{2j}} (2\theta_{A_1\to M_{1j}}+\theta_{A_1\to M_{2j}}) , \\
& \ldots & \\
\eta_j^{(T)} & = &  \theta_{M_{1j}\to R_1} \sum_{t=1}^T(T-t+1) \theta_{A_1\to M_{tj}}  \\
& & \hspace{1.5in} + \sum_{t=1}^{T-1} \theta_{M_{1j}\to R_{t+1}}^{M_{1j},\ldots,M_{(t+1)j}} \left\{\sum_{s=1}^{T-t} (T-t-s + 1) \theta_{A_1\to M_{sj}} \right\}.
\end{eqnarray*}

Dividing the right-hand side of the above equation by $T$ and taking $T\to\infty$, we obtain the formula of individual mediation effect in an infinite-horizon setting:
\[
\begin{split}
\eta_j^{(\infty)} 
& = \theta_{M_{1j}\to R_1} \underset{T\to\infty}{\lim}\frac{1}{T}\sum_{t=1}^{T}(T-t+1)\theta_{A_1\to M_{tj}} \\ 
& \hspace{1.5in} + \underset{T\to\infty}{\lim}\frac{1}{T} \sum_{t=1}^{T-1} \theta_{M_{1j}\to R_{t+1}}^{M_{1j},\ldots,M_{(t+1)j}} \left\{ \sum_{s=1}^{T-t} (T-t- s+1) \theta_{A_1\to M_{sj}} \right\} \\
& \equiv \theta_{M_{1j}\to R_1} I_{1j} + I_{2j}.
\end{split}
\]
We next derive the two terms $I_{1j}$ and $I_{2j}$, respectively, 

For $I_{1j}$, we have
\vspace{-0.01in}
\begin{equation*}\label{pf:expanded_I1}
	\begin{split}
	I_{1j} &= \lim_{T\to\infty} \frac{1}{T} \sum_{t=1}^{T}(T-t+1)\theta_{A_1\to M_{tj}} \\
	& = \lim_{T\to\infty} \frac{1}{T} \left\{ T\theta_{A_1\to M_{1j}} + (T-1) \theta_{A_1\to M_{2j}} +\ldots + \theta_{A_1\to M_{Tj}} \right\} \\ 
	& = \lim_{T\to\infty} \frac{1}{T} \sum_{t=1}^{T} \sum_{s=1}^{t} \theta_{A_1\to M_{sj}}.
	\end{split}
\end{equation*}

Following the recursive relation in \eqref{eq:recursion_total_effects} of the paper, we first obtain the following equations with the time-invariant model parameters,
\[
\begin{split}
\btheta_{A_1\to \bM_1} & = \bm{\delta}_1, \\
\btheta_{A_1\to \bM_2} & = \bGamma_{1} \btheta_{A_1\to \bM_1} + \bzeta_{1} \theta_{A_1\to R_1}, \\
&\ldots \\
\btheta_{A_1\to \bM_t} &= \bGamma_{1} \btheta_{A_1\to \bM_{t-1}} + \bzeta_{1} \theta_{A_1\to R_{t-1}}. 
\end{split}
\]
Taking summation over the above equations leads to
\begin{equation*}
	\sum_{s=1}^{t} \btheta_{A_1\to \bM_s} = \bdelta_1 + \bGamma_{1} \sum_{s=1}^{t-1} \btheta_{A_1\to \bM_s} + \bzeta_1 \sum_{s=1}^{t-1} \theta_{A_1\to R_s}.
\end{equation*}
Taking another summation of the above equations leads to
\begin{equation}\label{pf:recursion_A}
	\frac{1}{T}\sum_{t=2}^{T} \sum_{s=1}^{t} \btheta_{A_1\to \bM_s} =\frac{T-1}{T} \bdelta_1 + \bGamma_1 \frac{1}{T}\sum_{t=1}^{T-1}\sum_{s=1}^{t} \btheta_{A_1\to \bM_s} + \bzeta_1 \frac{1}{T} \sum_{t=1}^{T-1} \sum_{s=1}^{t}\theta_{A_1\to R_s}.
\end{equation}

Similarly, to evaluate $\sum_{s=1}^{t}\theta_{A_1\to R_s}$, we first obtain the recursion relations,
\[
\begin{split}
\theta_{A_1\to R_1} &=\delta_2 + \bkappa\trans\btheta_{A_1\to \bM_1}, \\
\theta_{A_1\to R_2} & = \zeta_2\theta_{A_1\to R_1} + \bkappa\trans\btheta_{A_1\to \bM_2}+ \bgamma_2\trans\btheta_{A_1\to \bM_1}, \\
&\ldots \\
\theta_{A_1\to R_t} & = \zeta_2\theta_{A_1\to R_{t-1}} + \bkappa\trans\btheta_{A_1\to \bM_t} + \bgamma_2\trans\btheta_{A_1\to \bM_{t-1}}.
\end{split}
\]
Taking double summation over the above equations leads to
\begin{eqnarray}\label{pf:recursion_R}
\frac{1}{T}\sum_{t=2}^{T}\sum_{s=1}^{t} \theta_{A_1\to R_s} & = & \frac{T-1}{T}\delta_2 + \zeta_2\frac{1}{T} \sum_{t=1}^{T-1}\sum_{s=1}^{t} \theta_{A_1\to R_s} \nonumber \\
& & \hspace{0.3in} + \bkappa\trans \frac{1}{T}\sum_{t=1}^{T-1}\sum_{s=1}^{t} \btheta_{A_1\to \bM_s} + \bgamma_{2}\trans \frac{1}{T}\sum_{t=1}^{T-1}\sum_{s=1}^{t}\btheta_{A_1\to \bM_s}
\end{eqnarray}

Denote ${\bB}_1\equiv \lim_{T\to\infty}T^{-1}\sum_{t=2}^{T}\sum_{s=1}^{t} \btheta_{A_1\to \bM_s}\in\mathbb{R}^d$, and $\widetilde{R}\equiv\lim_{T\to\infty}T^{-1} \sum_{t=2}^{T}$ $\sum_{s=1}^{t}\theta_{A_1\to R_s}$. Combining \eqref{pf:recursion_A} and~\eqref{pf:recursion_R}, we obtain that
\[
\begin{split}
\bB_1 & = \bm{\delta}_1 + \bGamma_{1}\bB_1 + \bzeta_1 \widetilde{R}, \\
\widetilde{R} &= \delta_2 + \zeta_2 \widetilde{R} + \bkappa\trans \bB_1 + \bgamma_2\trans \bB_1.
\end{split}
\]
Solving the above equations leads to
\[
\bB_1 = \left\{(1-\zeta_2)(\bI - \bGamma_{1})-\bzeta_1(\bkappa+\bgamma_{2})\trans\right\}^{-1} \left\{(1-\zeta_2)\bm{\delta}_1 + \delta_2 \bzeta_1 \right\}.
\]
Then $I_{1j}$ is the $j$th element in the vector ${\bB}_1$.

For $I_{2j}$, we have 
\[
\begin{split}
I_{2j} &=\underset{T\to\infty}{\lim}\frac{1}{T} \sum_{t=1}^{T-1} \theta_{M_{1j}\to R_{t+1}}^{M_{1j},\ldots,M_{(t+1)j}} \left\{ \sum_{s=1}^{T-t} (T-t- s+1) \theta_{A_1\to M_{sj}}  \right\} \\
&= \underset{T\to\infty}{\lim}\frac{1}{T} \sum_{t=1}^{T-1} \theta_{M_{1j}\to R_{t+1}}^{M_{1j},\ldots,M_{(t+1)j}} \left(\sum_{s=1}^{T-t} \sum_{i=1}^{s} \theta_{A_1\to M_{ij}} \right).
\end{split}
\]

For notational simplicity, let $L_{tj}\equiv\sum_{s=1}^{t}\sum_{i=1}^{s} \theta_{A_1\to M_{ij}}$. Next, we derive the recursive relationship for the summation $T^{-1}\sum_{t=1}^{T-1}\theta_{M_{1j}\to R_{t+1}}^{M_{1j},\ldots,M_{(t+1)j}}L_{(T-t)j}$. 

Following the recursion for $\theta_{M_{sj}\to R_{t}}^{M_{sj},\ldots,M_{tj}}$ in \eqref{eq:finite_single_term_M}, and multiplying both sides by $L_{tj}$, we obtain that
\[
\begin{split}
 \theta_{M_{1j}\to R_2}^{M_{1j},M_{2j}} L_{(T-1)j} & =L_{(T-1)j} \theta_{M_{1j}\to R_2} -  \theta_{M_{1j}\to M_{2j}} \theta_{M_{1j}\to R_1} L_{(T-1)j}, \\
 \theta_{M_{1j}\to R_3}^{M_{1j},\ldots,M_{3j}} L_{(T-2)j} &= L_{(T-2)j}  \theta_{M_{1j}\to R_3} - \theta_{M_{1j}\to M_{3j}} \theta_{M_{1j}\to R_1} L_{(T-2)j}- 
\theta_{M_{1j}\to M_{2j}} \theta_{M_{1j}\to R_{2}}^{M_{1j}, M_{2j}} L_{(T-2)j}, \\
\theta_{M_{1j}\to R_4}^{M_{1j},\ldots,M_{4j}}L_{(T-3)j} &=L_{(T-3)j} \theta_{M_{1j}\to R_4} 
-\theta_{M_{1j}\to M_{4j}} \theta_{M_{1j}\to R_1} L_{(T-3)}
- \sum_{i=2}^{3}\theta_{M_{1j}\to M_{ij}} \theta_{M_{1j}\to R_{4-i+1}}^{M_{1j},\ldots,M_{(4-i+1)j}} L_{(T-3)j}, \\
&\ldots \\
 \theta_{M_{1j}\to R_{T}}^{M_{1j},\ldots,M_{Tj}} L_{1j}&= L_{1j} \theta_{M_{1j}\to R_T} - \theta_{M_{1j}\to M_{Tj}}\theta_{M_{1j}\to R_1}L_{1j} - \sum_{i=2}^{T-1} \theta_{M_{1j}\to M_{ij}} \theta_{M_{1j}\to R_{T-i+1}}^{M_{1j},\ldots,M_{(T-i+1)j}}  L_{1j}.
\end{split}
\]
Taking summation over the above equations leads to 
\[
\begin{split}
\frac{1}{T}\sum_{t=1}^{T-1} \theta_{M_{1j}\to R_{t+1}}^{M_{1j},\ldots,M_{(t+1)j}}  L_{(T-t)j} &= 
\frac{1}{T} \sum_{t=1}^{T-1}  \theta_{M_{1j}\to R_{t}}L_{(T-t)j} - \theta_{M_{1j}\to R_1}\frac{1}{T}\sum_{t=1}^{T-1}\theta_{M_{1j}\to M_{tj}}L_{(T-t)j} \\
&- \sum_{i=2}^{T-1} \theta_{M_{1j}\to M_{ij}} \frac{1}{T} \sum_{t=1}^{T-i} \theta_{M_{1j}\to R_{t+1}}^{M_{1j},\ldots,M_{(t+1)j}} L_{(T-i-t+1)j}.
\end{split}
\]
Define ${\bB}_2 \equiv \lim_{T\to\infty}\sum_{t=2}^{T}\btheta_{M_{1j}\to \bM_t}\in\mathbb{R}^d$, $B_{3j} \equiv \lim_{T\to\infty}\sum_{t=2}^{T}\theta_{M_{1j}\to R_t}$, ${\bB}_4 \equiv \lim_{T\to\infty}$ $T^{-1} \sum_{t=2}^{T}$ $\btheta_{M_{1j}\to \bM_t} L_{(T-t+1)j}$, and $B_{5j} \equiv \lim_{T\to\infty}T^{-1}\sum_{t=2}^{T} \theta_{M_{1j}\to R_{t}} L_{(T-t+1)j}$.  Taking the limit with $T\to\infty$ on both sides of the above equation leads to
\begin{equation}\label{eq:I_2}
I_{2j} =(1+B_{2j})^{-1} (B_{5j} - \theta_{M_{1j}\to R_1}B_{4j}).
\end{equation}

Next, we obtain the recursive relationship for $\sum_{t=2}^{T}\btheta_{M_{1j}\to \bM_{t}}$, $\sum_{t=2}^{T}\btheta_{M_{1j}\to\bM_t} L_{(T-t+1)j}$, and $\sum_{t=2}^{T}\theta_{M_{1j}\to R_t} L_{(T-t+1)j}$. Let $\bB_6 \equiv \begin{pmatrix} \bzero_{d\times d} & \bzero_{d} \\ \bkappa\trans & 0 \end{pmatrix}$ and $\bB_7 \equiv \begin{pmatrix} \bGamma_1 & \bzeta_1 \\ \bgamma_2\trans & \zeta_2 \end{pmatrix}$. The recursion between $(\btheta_{M_{1j}\to \bM_t},\theta_{M_{1j}\to R_t})$ and $(\btheta_{M_{1j}\to \bM_{t-1}}, \theta_{M_{1j}\to R_{t-1}})$ takes the following form:
\begin{align}\label{eq:recursion_B_C}
\begin{bmatrix} \btheta_{M_{1j}\to \bM_t} \\ \theta_{M_{1j}\to R_t}\end{bmatrix} 
= \begin{pmatrix} \bzero_{d\times d} & \bzero_{d} \\ \bkappa\trans & 0 \end{pmatrix} \begin{bmatrix} \btheta_{M_{1j}\to \bM_t} \\ \theta_{M_{1j}\to R_t} \end{bmatrix} + \begin{pmatrix} \bGamma_1 & \bzeta_1 \\ \bgamma_2\trans & \zeta_2 \end{pmatrix} \begin{bmatrix} \btheta_{M_{1j}\to \bM_{t-1}} \\ \theta_{M_{1j}\to R_{t-1}}\end{bmatrix}
\end{align}
Taking summation from $t=2$ to $T$ of the above equation leads to
\begin{align*}
\sum_{t=2}^T (\bI-\bB_6) \begin{bmatrix} \btheta_{M_{1j}\to \bM_t} \\ \theta_{M_{1j}\to R_{t}}\end{bmatrix} = \sum_{t=2}^T \bB_7 \begin{bmatrix} \btheta_{M_{1j}\to \bM_{t-1}} \\ \theta_{M_{1j}\to R_{t-1}} \end{bmatrix} \\
(\bI-\bB_6)\sum_{t=2}^T  \begin{bmatrix} \btheta_{M_{1j}\to \bM_t} \\ \theta_{M_{1j}\to R_{t}} \end{bmatrix} = \bB_7 \sum_{t=2}^{T-1} \begin{bmatrix} \btheta_{M_{1j}\to \bM_t} \\ \theta_{M_{1j}\to R_{t}} \end{bmatrix} + \bB_7 \begin{bmatrix} \btheta_{M_{1j}\to \bM_1} \\ \theta_{M_{1j}\to R_{1}} \end{bmatrix}
\end{align*}
Taking the limit with $T\to \infty$ on both sides leads to
\begin{align}\label{eq:B_C_bar}
\begin{bmatrix} \bB_2 \\ B_{3j}\end{bmatrix} = (\bI-\bB_6-\bB_7)^{-1}\bB_7 \begin{bmatrix} \btheta_{M_{1j}\to \bM_1} \\ \theta_{M_{1j}\to R_1}\end{bmatrix}.
\end{align}

Multiply both sides of the recursion in \eqref{eq:recursion_B_C} by $L_{(T-t+1)j}$ leads to
\begin{align*}
& \begin{bmatrix} \btheta_{M_{1j}\to \bM_t} L_{(T-t+1)j} \\ \theta_{M_{1j}\to R_{t}} L_{(T-t+1)j}\end{bmatrix} = \bB_6 \begin{bmatrix} \btheta_{M_{1j}\to \bM_t} L_{(T-t+1)j} \\ \theta_{M_{1j}\to R_{t}} L_{(T-t+1)j} \end{bmatrix} + \bB_7 \begin{bmatrix} \btheta_{M_{1j}\to \bM_{t-1}} L_{(T-t+1)j} \\ \theta_{M_{1j}\to R_{t-1}} L_{(T-t+1)j} \end{bmatrix},\\
& \frac{1}{T}\sum_{t=2}^T (\bI-\bB_6) \begin{bmatrix} \btheta_{M_{1j}\to \bM_t} L_{(T-t+1)j} \\ \theta_{M_{1j}\to R_{t}} L_{(T-t+1)j}\end{bmatrix} = \frac{1}{T}\sum_{t=2}^T \bB_7 \begin{bmatrix} \btheta_{M_{1j}\to \bM_{t-1}}L_{(T-t+1)j} \\ \theta_{M_{1j}\to R_{t-1}}L_{(T-t+1)j} \end{bmatrix} \\
& (\bI-\bB_6-\bB_7) \frac{1}{T}\sum_{t=2}^T  \begin{bmatrix} \btheta_{M_{1j}\to \bM_t} L_{(T-t+1)j} \\ \theta_{M_{1j}\to R_{t}} L_{(T-t+1)j}\end{bmatrix} =  \frac{1}{T}\bB_7 \begin{bmatrix} \btheta_{M_{1j}\to \bM_1} L_{(T-1)j} \\ \theta_{M_{1j}\to R_1} L_{(T-1)j}\end{bmatrix}
\end{align*}

Furthermore, we have $\lim_{T\to\infty}\frac{1}{T}L_{(T-1)j}=\lim_{T\to\infty}\frac{1}{T}\sum_{t=1}^{T-1}\sum_{i=1}^t\theta_{A_1\to M_{ij}}=B_{1j}$. We now take the limit with $T\to\infty$ on both sides of the summations, we have
\begin{align}\label{eq:B_C_tilde}
\begin{bmatrix} \bB_4 \\ B_{5j} \end{bmatrix} = (\bI-\bB_6-\bB_7)^{-1}\bB_7 \begin{bmatrix} \btheta_{M_{1j}\to \bM_1} \\ \theta_{M_{1j}\to R_1}\end{bmatrix} B_{1j}.
\end{align}
We then combine \eqref{eq:I_2}, \eqref{eq:B_C_bar} and \eqref{eq:B_C_tilde} and solve for $I_{2j}$.

This completes the proof of Theorem \ref{thm:eta_infinite}.
\eop

\subsection{Proof of Theorem \ref{theory:finite}}

We begin with some notations. Let $\text{vech}(\bB) \in \mathbb{R}^{q(q+1)/2}$ denote the half-vectorization of a symmetric $q\times q$ matrix $\bB$ that vectorizes the lower triangular part of $\bB$. 
Let ${\partial \by}/{\partial \bx}$ denote the $r \times s$ matrix of the derivative of $\by(\bx) = \left(y_1(\bx),\ldots,y_r(\bx)\right)\trans \in \Rbb^{r}$ with respect to $\bx=(x_1,\ldots,x_s)\trans \in \Rbb^{s}$ whose $(i,j)$th entry is ${\partial y_i}/{\partial x_j}, i=1,\ldots,r, j=1,\ldots,s$. 

We first prove the case of $T=1$, then the case of $T=2$. The case for any finite $T > 2$ can be proved in a similar fashion. 

For $T=1$, recall $\eta_j^{(1)} = \theta_{A_1\to M_{1j}}\theta_{M_{1j}\to R_1} = \beta_{1j}\beta_{M_{1j},R_1\mid\Pa_j\cup A_1}$, and $\widehat{\eta}_j^{(1)} = \widehat{\theta}_{A_1\to M_{1j}}$ $\widehat{\theta}_{M_{1j}\to R_1}=\widehat{\beta}_{1j}\widehat{\beta}_{M_{1j},R_1\mid \Pa_j\cup A_1}$ as $n\to\infty$. To establish the asymptotic distribution of $\widehat{\eta}_j^{(1)}$, the main idea is to first show the asymptotic distribution of $(\widehat{\theta}_{A_1\to M_{1j}}, \widehat{\theta}_{M_{1j}\to R_1})\trans$, then apply the delta method.

Let $\bU_{1j} \equiv(U_{1j1},\ldots,U_{1jq_1})\trans=(A_1,\Pa_{1j}\trans,M_{1j},R_1)\trans \in \Rbb^{q_1}$. Let $\bSigma_{1j} = \text{cov}(\bU_{1j})\in\mathbb{R}^{q_1\times q_1}$, and let $\widehat{\bSigma}_{1j}$ denote the corresponding sample covariance matrix. Since all fourth moments of the variables in $\bU_{1j}$ are finite, the central limit theorem implies that
\begin{equation}\label{eq:MVN_theta}
\sqrt{n}\left\{ \text{vech}(\widehat{\bSigma}_{1j}) - \text{vech}(\bSigma_{1j}) \right\} \overset{d}{\to} \mathcal{N}(\bm{0}, \bV_1),
\end{equation} 
where $\bV_{1j} = \text{cov}\left(\text{vech}(\bU_{1j}\bU_{1j}\trans) \right)$. 

Let $\btheta_{1j} \equiv (\theta_{A_1\to M_{1j}}, \theta_{M_{1j}\to R_1})\trans = (\beta_{1j}, \beta_{M_{1j},R_1\mid \Pa_{1j}\cup A_1})\trans$, and $\widehat{\btheta}_{1j} \equiv (\widehat{\theta}_{A_1\to M_{1j}}, \widehat{\theta}_{M_{1j}\to R_1})\trans$ $= (\widehat{\beta}_{1j}, \widehat{\beta}_{M_{1j}, R_1\mid \Pa_{1j}\cup A_1})\trans$. There exists a differentiable function $h_{11}:\mathbb{R}^{q_1(q_1+1)/2}\to\mathbb{R}^2$, such that $h_{11}(\vh(\bSigma_{1j}))=\btheta_{1j}$, and $h_{11}(\vh(\widehat{\bSigma}_{1j}))=\widehat{\btheta}_{1j}$. Therefore, by \eqref{eq:MVN_theta} and the multivariate delta method, we have
\[
\sqrt{n}(\widehat{\btheta}_{1j}- \btheta_{1j}) \overset{d}{\to} \mathcal{N}(\bm{0}, \bLambda_{1j}\bV_{1j}\bLambda_{1j}\trans),
\]
where $\bLambda_{1j}=\partial \btheta_{1j}/\partial \vh(\bSigma_{1j})$. 

Consider another differentiable function $h_{12}(x_1,x_2)=x_1x_2$, and apply the multivariate delta method again, we have 
\[
\sqrt{n}(\widehat{\eta}_j^{(1)} - \eta_j^{(1)}) \overset{d}{\to}\mathcal{N}(0, \bS_{1j}\bLambda_{1j}\bV_{1j}\bLambda_{1j}\trans\bS_{1j}\trans),
\]
where $\bV_{1j} = \text{cov}\left(\text{vech}(\bU_{1j}\bU_{1j}\trans) \right)$, $\bLambda_{1j}=\partial \btheta_{1j}/\partial \left(\vh(\bSigma_{1j})\right)$, $\btheta_{1j}=(\theta_{A_1\to M_{1j}},\theta_{M_{1j}\to R_1})\trans$, and $\bS_{1j} = {\partial \eta_j^{(1)}}/{\partial \btheta_{1j}}$.

For $T=2$, let $\bU_{2j} \equiv(U_{2j1},\ldots, U_{2jq_2})\trans=(A_1, \bM_1\trans, R_1, A_2, \Pa_{2j}\trans, M_{2j}, R_2)\trans \in \Rbb^{q_2}$, where $\Pa_{2j} = \{M_{2i}:(\bW_t)_{ij}\neq 0,i\in\{1,\ldots,d \}\setminus \{j\} \}$ is the parent set of $M_{2j}$ in $\mathcal{G}_2$. Let $\bSigma_{2j} = \text{cov}(\bU_{2j})\in\mathbb{R}^{q_2\times q_2}$ and let $\widehat{\bSigma}_{2j}$ denote the corresponding sample covariance matrix. Similar to \eqref{eq:MVN_theta}, we have
\[
\sqrt{n}\left\{ \text{vech}(\widehat{\bSigma}_{2j}) - \text{vech}(\bSigma_{2j}) \right\} \overset{d}{\to} \mathcal{N}(\bm{0}, \bV_{2j}),
\]
where $\bV_{2j} = \text{cov}\left(\text{vech}(\bU_{2j}\bU_{2j}\trans) \right)$. 

Let $\btheta_{2j}\equiv(\theta_{A_1\to M_{1j}}, \theta_{M_{1j}\to R_1}, \theta_{A_1\to M_{2j}}, \theta_{M_{1j}\to {M_{2j}}}, \theta_{M_{1j}\to R_2}, \theta_{A_2\to M_{2j}}, \theta_{M_{2j}\to R_2})\trans$, and $\widehat{\btheta}_{2j} \equiv (\widehat{\theta}_{A_1\to M_{1j}},$  $\widehat{\theta}_{M_{1j}\to R_1}, \widehat{\theta}_{A_1\to M_{2j}}, \widehat{\theta}_{M_{1j}\to {M_{2j}}}, \widehat{\theta}_{M_{1j}\to R_2}, \widehat{\theta}_{A_2\to M_{2j}}, \widehat{\theta}_{M_{2j}\to R_2})\trans$.  Note that every element in $\btheta_{2j}$ represents the total effect of one variable on another, where all variables in this expanded DAG up to the second stage are included in $\bU_{2j}$. For example, $\theta_{M_{1j}\to R_2}=\beta_{M_{1j},R_2\mid\Pa_j\cup A_1}$ where this regression coefficient depends only on the covariance matrix of $\bU_{2j}$, which is $\bSigma_{2j}$. Then there exists a differentiable function $h_{21}:\mathbb{R}^{q_2(q_2+1)/2}\to \mathbb{R}^7$, such that $h_{21}(\vh(\bSigma_{2j}))=\btheta_{2j}$ and $h_{21}(\vh(\widehat{\bSigma}_{2j}) )=\widehat{\btheta}_{2j}$. By the multivariate delta method, we have
\[
\sqrt{n}(\widehat{\btheta}_{2j}- \btheta_{2j}) \overset{d}{\to} \mathcal{N}(\bm{0}, \bLambda_{2j}\bV_{2j}\bLambda_{2j}\trans),
\]
where $\bLambda_{2j} = {\partial\btheta_{2j}}/{\partial \left(\vh(\bSigma_{2j})\right)}$. 

Consider another differentiable function $h(x_1,x_2,\ldots,x_7)=\{ x_1x_2 + (x_3 + x_6)\cdot x_7 + x_1(x_5 - x_4 x_7)\} /2 $, and apply the multivariate delta method again, we have
\[
\sqrt{n}(\widehat{\eta}_j^{(2)} - \eta_j^{(2)}) \overset{d}{\to}\mathcal{N}(0, \bS_{2j}\bLambda_{2j}\bV_{2j}\bLambda_{2j}\trans\bS_{2j}\trans),
\]
where $\bV_{2j} = \text{cov}\left(\text{vech}(\bU_{2j}\bU_{2j}\trans) \right)$, $\bLambda_{2j} = \partial \btheta_{2j}/\partial \left(\vh(\bSigma_{2j})\right)$, and $\bS_{2j} = {\partial \eta_j^{(2)}}/{\partial \btheta_{2j}}$. 

Generally, for $t=1,\ldots, T$, let $\bU_{tj} \equiv (U_{tj1}, \ldots, U_{tjq_t})\trans = (\bar{A}_{t-1}, \bar{\bM}_{t-1}\trans, \bar{R}_{t-1}, A_t, \Pa_{tj}\trans, M_{tj}, R_t)\trans \in \Rbb^{q_t}$, where $\Pa_{tj} = \{M_{ti}:(\bW_{t})_{ij}\neq 0,i\in\{1,\ldots,d \}\setminus \{j\} \}$ is the parent set of $M_{tj}$ in $\mathcal{G}_t$, and $\btheta_{tj} \equiv (\btheta_{(t-1)j}\trans, \theta_{A_1\to M_{tj}},\ldots,\theta_{A_{t}\to M_{tj}},\theta_{M_{1j}\to {M_{tj}}}, \ldots, \theta_{M_{t-1,j}\to {M_{tj}}}, \theta_{M_{1j}\to R_t}, \ldots, \theta_{M_{tj}\to R_t})\trans$ $\in \Rbb^{p_t}$.
For $T>2$, it follows that 
\begin{equation*}
	\sqrt{n}(\widehat{\eta}_j^{(t)} - \eta_{j}^{(t)}) \overset{d}{\to}\mathcal{N}(0, \sigma_{tj}^2), \ \text{as} \ n \to \infty, 
	\vspace{-0.01in}
\end{equation*}
where $\sigma_{tj}^2=\bS_{tj}\bLambda_{tj}\bV_{tj}\bLambda_{tj}\trans\bS_{tj}\trans$, $\bV_{tj} = \text{cov}\left(\text{vech}(\bU_{tj}\bU_{tj}\trans) \right)\in\mathbb{R}^{q_t(q_t+1)/2\times q_t(q_t+1)/2}$, $\bLambda_{tj}= \partial \btheta_{tj}/\partial \left(\vh(\bSigma_{tj})\right)\in\mathbb{R}^{p_t\times q_t(q_t+1)/2}$, $\bSigma_{tj} = \text{cov}(\bU_{tj})\in\Rbb^{q_t\times q_t}$, and $\bS_{tj} = {\partial {\eta_j^{(t)}}} / \partial \btheta_{tj}\in\mathbb{R}^{1\times p_t}$.

This completes the proof of Theorem \ref{theory:finite}. 
\eop

\subsection{Proof of Theorem \ref{theory:infinite}}

From the proof of Theorem \ref{thm:eta_infinite}, recall that $I_{1j} = B_{1j}$, and $I_{2j} = (1+ {B}_{2j})^{-1}(B_{5j}- \theta_{M_{1j}\to R_1}B_{4j})$. Therefore, $I_{1j}$, $I_{2j}$ are continuous functions of the intermediate quantities $(\theta_{M_{1j}\to R_1},\btheta_{M_{1j}\to \bM_1})$, and the model parameters $\bTheta_1=(\balpha_1, \bbeta_1, \bGamma_{1}, \bzeta_1)\trans\in\mathbb{R}^{(d+3)\times d}$, $\bTheta_2=(\alpha_2, \beta_2, \zeta_2, \bgamma_2\trans, \bkappa\trans)\trans\in\mathbb{R}^{(2d+3)\times 1}$. To establish the asymptotic distribution of $\widehat{\eta}_j^{(\infty)}=\widehat{\theta}_{M_{1j}\to R_1}\widehat{I}_{1j}+\widehat{I}_{2j}$, the main idea is to first show the asymptotic distribution of the estimators of the model parameters $(\bTheta_1,\bTheta_2)$ and those intermediate quantities $(\bB_1,\bB_2, B_{3j}, \bB_4, B_{5j})$, then apply the delta method. 
		
We first show the asymptotic normality of $\widehat{\bTheta}_2$. The asymptotic normality of $\widehat{\bTheta}_1$ can be shown similarly. Let $\bX_{it}\equiv({1}, A_{it}, \bM_{i(t-1)}\trans, R_{i(t-1)},\bM_{it}\trans )\trans\in\mathbb{R}^{(2d+3)\times 1}$. Since 
\[
\sqrt{nT}(\widehat{\bTheta}_2 - \bTheta_2) =\left(\frac{1}{nT}\sum_{i=1}^{n}\sum_{t=1}^T\bX_t\bX_t\trans\right)^{-1} \left(\frac{1}{\sqrt{nT}}\sum_{i=1}^{n}\sum_{t=1}^T\bX_{it}\epsilon_{i,rt}\right),
\]
and the stochastic process $\{(A_t, \bM_t\trans, R_t):t\geq 1\}$ is $\beta$-mixing, by the weak law of large numbers of a weak mixing process, we have
\vspace{-0.01in}
\[
\begin{split}
&\frac{1}{nT}\sum_{i=1}^{n}\sum_{t=1}^{T} \bX_{it} \bX_{it}\trans \overset{p}{\to} \mathbb{E}[\bX_0\bX_0\trans], \ \text{where} \ \mathbb{E}[\bX_0\bX_0\trans] = \lim_{nT\to\infty}\frac{1}{nT}\sum_{i=1}^{n}\sum_{t=1}^{T}\mathbb{E}[\bX_{it}\bX_{it}\trans],\\
&\frac{1}{nT}\sum_{i=1}^{n}\sum_{t=1}^{T} \bX_{it} \epsilon_{i,rt} \overset{p}{\to} \bm{0}, \text{because} \ \bX_{it} \indep \epsilon_{i,rt}, \ \text{and} \ \mathbb{E}[\epsilon_{i,rt}] = {0}, \ i=1,\dots,n.
\end{split}
\]

Next, we derive the asymptotic distribution of $\widetilde{\bU}_T=\sum_{t=1}^T \bU_t \equiv\sum_{t=1}^T \bX_t\epsilon_{rt}\in\mathbb{R}^{(2d+3)\times 1}$. For notional simplicity, we drop the subscript $i$ in this part. To apply the Cram{\'e}r-Wold theorem, let $\bm{a}=(a_1,\ldots,a_{2d+3})\trans\in\mathbb{R}^{2d+3}$ be a nonrandom and nonzero vector, and $\bU_t = (U_{t1},\ldots,U_{t,2d+3})\trans$. Then we write
\begin{equation*}\label{eq:Cramer-Wold}
\bm{a}\trans\widetilde{\bU}_T = \sum_{t=1}^T\sum_{p=1}^{2d+3} a_p U_{tp} = \sum_{t=1}^T U_t^\prime.
\end{equation*}
Since $\{\bU_t\}_{t\geq 1}$ is a $\beta$-mixing process with mean zero, it follows that $\{U_t^\prime\}_{t\geq 1}$ is also a $\beta$-mixing centered process with the second moment given by
\[
  \sigma^2_T = \bm{a}\trans\text{var}(\widetilde{\bU}_T)\bm{a}= \text{var}\left(\sum_{t=1}^T U_t^\prime\right) \to \infty, \ T\to\infty.
\]
For every $\epsilon>0$,
\[
\begin{split}
\sum_{t=1}^T \mathbb{E}\left[(U_t^\prime)^2\mathbf{1}(|U_t^\prime|>\epsilon\sigma_T)\right] &= \sum_{t=1}^T\mathbb{E}
\left\{\left(\sum_{p=1}^{2d+3} a_pU_{tp}\right)^2 \mathbf{1}\left( \sum_{p=1}^{2d+3}|a_pU_{tp} | > \epsilon\sigma_T\right) \right\} \\
& \leq \sum_{t=1}^T a_p^2 \mathbb{E}\left\{ U_{tp}^2 \mathbf{1}\left(\sum_{p=1}^{2d+3} |U_{tp}|>\frac{\epsilon\sigma_T}{\max_p|a_p|} \right) \right\},
\end{split}
\]
where $\mathbf{1}(\cdot)$ is an indicator function. Since $\sigma_T\to\infty$, and $\max_p|a_p|<\infty$, we have 
\[
\mathbf{1}\left(\sum_{p=1}^{2d+3} |U_{tp}|>\frac{\epsilon\sigma_T}{\max_p|a_p|} \right) \overset{a.s.}{\to}0. 
\]
In addition, because $\mathbb{E}[U_{tp}^2]<\infty$, and $\mathbb{P}(U_{tp}=\infty)=0$, it follows that
\[
\sum_{t=1}^T \mathbb{E}\left[(U_t^\prime)^2\mathbf{1}(|U_t^\prime|>\epsilon\sigma_T)\right] \to 0, \ T \to\infty.
\]
By the central limit theorem for martingales, we have
\[
  \frac{\sum_{t=1}^T U_t^\prime}{\sigma_T}\overset{d}{\to}\mathcal{N}(0,1), \ \text{as} \ T\to\infty.
\]
Applying the Cram{\'e}r-Wold theorem, we have
\[
\begin{split}
  \frac{1}{\sqrt{T}}\sum_{t=1}^T\bU_t=\frac{1}{\sqrt{T}}\sum_{t=1}^T \bX_t \epsilon_{rt} \overset{d}{\to}\mathcal{N}(\bm{0}, \bV_{r}),
\end{split}
\]
where $\bV_r = \sigma_r^2\mathbb{E}(\bX_t\bX_t\trans)\in\mathbb{R}^{(2d+3)\times(2d+3)}$. Generalizing to $n$ i.i.d. subjects and by Slutsky's theorem, 
\[
\begin{split}
\sqrt{nT}(\widehat{\bTheta}_2 - \bTheta_2) = \left(\frac{1}{nT}\sum_{i=1}^{n}\sum_{t=1}^{T}\bX_{it} \bX_{it}\trans\right)^{-1}\left(\frac{1}{\sqrt{nT}}\sum_{i=1}^{n}\sum_{t=1}^{T}\bX_{it} \epsilon_{i,rt}\right)\overset{d}{\to}\mathcal{N}(\bm{0}, \bV_{\theta_2}), 
\end{split}
\]
where $\bV_{\theta_2}=\mathbb{E}[\bX_0\bX_0\trans]^{-1}\bV_{r}\mathbb{E}[\bX_0\trans\bX_0]^{-1}\in\mathbb{R}^{(2d+3)\times(2d+3)}$. 

{Since $\bB_1=\left\{(1-\zeta_2)(\bI - \bGamma_1) - \bzeta_1(\bkappa + \bgamma_2)\trans \right\}^{-1}\left\{(1-\zeta_2)\bdelta_1 + \delta_2\bzeta_1 \right\}$ is a continuous function of $(\bTheta_1,\bTheta_2)$, then by multivariate delta method, the estimator of $\bB_1$ which is a continuous function of $(\widehat{\bTheta}_1,\widehat{\bTheta}_2)$ is also asymptotically normal. Therefore, we establish the asymptotic normality of $\widehat{I}_{1j}=\widehat{B}_{1j}$.}
		
Next, we show the asymptotic normality of $\widehat{\theta}_{M_{1j}\to R_1}$. Let $\bY_t\equiv(\bM_{t-1}\trans, R_{t-1}, A_t, \Pa_j, M_{tj},$ $R_t)\trans\in\mathbb{R}^{q\times 1}$, and $\mathbb{E}(\bY_t)=\bmu_Y$ is independent of $t$ because of the stationarity assumption. Let ${\bSigma} = \mathbb{E}[(\bY_t-\bmu_Y)(\bY_t-\bmu_Y)\trans]$, and $\widehat{\bSigma} = T^{-1} \sum_{t=1}^{T}(\bY_t - \bar{\bY})(\bY_t -\bar{\bY})\trans$, where $\bar{\bY} = T^{-1} \sum_{t=1}^T\bY_t$. Since $\{\bY_t - \bmu_Y\}_{t\geq 1}$ is a centered stationary process with mean zero, by the finite fourth moments condition on $\bY_t$'s and the Martingale central limit theorem, we have
\[
\sqrt{nT}\left\{ \vh(\widehat{\bSigma}) - \vh(\bSigma) \right\} \overset{d}{\to}\mathcal{N}(\bm{0}, \bV_0),
\]
where $\bV_0 = \text{cov}\left[\vh\left\{(\bY_t-\bmu_Y)(\bY_t-\bmu_Y)\trans\right\} \right]\in\mathbb{R}^{q(q+1)/2\times q(q+1)/2}$.
		
Since $\theta_{M_{1j}\to R_1}=\ldots=\theta_{M_{tj}\to R_t}=\beta_{M_{tj},R_t\mid\Pa_j\cup \bM_{t-1}\cup R_{t-1}\cup A_t}$ for $t=2,\ldots,T$, there exists a differentiable function $h(\cdot):\mathbb{R}^{q(q+1)/2}\to \mathbb{R}$, such that $h(\vh(\bSigma))=\theta_{M_{1j}\to R_1}$, and $h(\vh(\widehat{\bSigma}))=\widehat{\theta}_{M_{1j}\to R_1}$. Then by the multivariate delta method, we have
\[
\sqrt{nT}(\widehat{\theta}_{M_{1j}\to R_1} - {\theta}_{M_{1j}\to R_1}) \overset{d}{\to}\mathcal{N}(0, \bLambda_0\bV_{0}\bLambda_0\trans),
\]
where $\bLambda_0 = \partial {\theta}_{M_{1j}\to R_1} / \partial\left(\vh(\bSigma)\right)\in\mathbb{R}^{1\times q(q+1)/2}$.

Similarly, we can establish the asymptotic normality of $\widehat{\btheta}_{M_{1j}\to \bM_1}$. Furthermore, since $({\bB}_2,B_{3j})$ and $({\bB}_4, {B}_{5j})$ are continuous functions of $(\bTheta_1,\bTheta_2,\btheta_{M_{1j}\to \bM_1}, \theta_{M_{1j}\to R_1})$, by the multivariate delta method again, the estimator of ${I}_{2j}=(1+B_{2j})^{-1}(B_{5j} - \theta_{M_{1j}\to R_1} {B}_{4j})$ is also asymptotically normal.

Furthermore, since we have established
\[
\sqrt{nT}\begin{pmatrix}
\widehat{\theta}_{M_{1j}\to R_1} -{\theta}_{M_{1j}\to R_1} \\
\widehat{I}_{1j} - I_{1j} \\
\widehat{I}_{2j} - I_{2j}
\end{pmatrix} \overset{d}{\to} \mathcal{N}(\bm{0}, \bSigma_j),
\]
where $\bSigma_j\in\mathbb{R}^{3\times 3}$ is a covariance matrix. Finally, since $\eta_j^{(\infty)}=\theta_{M_{1j}\to R_1} I_{1j}+I_{2j}$ and $\widehat{\eta}_j^{(\infty)} = \widehat{\theta}_{M_{1j}\to R_1}\widehat{I}_{1j}+\widehat{I}_{2j}$. 
Given a differentiable function $h(x_1,x_2,x_3)=x_1x_2 + x_3$, and by the multivariate delta method again, we have
\[
\sqrt{nT}(\widehat{\eta}_j^{(\infty)} - \eta_j^{(\infty)}) \overset{d}{\to}\mathcal{N}(0,\sigma_j^2), \ \text{as} \ nT\to\infty,
\]
where $\sigma_j^2 = \bV_j\trans\bSigma_j\bV_j\in\mathbb{R}$ where $\bV_j = (I_{1j}, \theta_{M_{1j}\to R_1}, 1)\trans$.
This completes the proof of Theorem \ref{theory:infinite}. 
\eop

\end{document}